\documentclass[pss]{wiley2sp} 
\usepackage{graphicx}
\usepackage{dcolumn}
\usepackage{amsmath}
\usepackage{amssymb}
\usepackage{verbatim}
\usepackage{color}
\usepackage{bm} 
\usepackage{bbm}
\usepackage{xspace}
\usepackage{marginnote}
\usepackage{mathtools}
\usepackage{dsfont}
\usepackage{hyperref} \hypersetup{colorlinks=true,linktoc=all,linkcolor=blue,breaklinks=true,citecolor=blue,urlcolor=blue}
\usepackage{etoolbox}
\usepackage[normalem]{ulem}

\robustify{\uparrow}
\robustify{\downarrow}
\robustify{\sum}
\robustify{\int}
\robustify{\nonumber}
\robustify{\cite}
\robustify{\footnote}
\newrobustcmd{\Figure}[2]{
  \begin{figure}[ht]
    \includegraphics[width=1.0\linewidth]{#1}
    \caption{#2}
  \end{figure}
}
\expandafter\newrobustcmd\csname Figure2\endcsname[3]{
\begin{figure}[ht]
  \includegraphics[width=0.9\linewidth]{#1}
  \\
  \includegraphics[width=0.9\linewidth]{#2}
  \caption{#3}
\end{figure}
}
\renewrobustcmd{\Re}{{\text{Re}}}
\renewrobustcmd{\Im}{{\text{Im}}}
\newrobustcmd{\eff}{\text{eff}} 
\newrobustcmd{\dagtot}{{\dag_\tot}}
\newrobustcmd{\dagres}{{\dag_\res}}
\newrobustcmd{\Ttot}{{\text{T}_\tot}}
\newrobustcmd{\Tres}{{\text{T}_\res}}
\newrobustcmd{\T}{{\text{T}}}
\newrobustcmd{\tot}{\text{tot}}
\newrobustcmd{\tun}{\text{tun}}
\newrobustcmd{\res}{\text{lead}}
\newrobustcmd{\un}{\text{i}}   
\newrobustcmd{\In}{\text{0}}   
\newrobustcmd{\state}{inverted stationary state\xspace}   
\newrobustcmd{\K}{\mathcal{K}}
\newrobustcmd{\D}{\mathcal{I}}
\newrobustcmd{\A}{\mathcal{A}}
\newrobustcmd{\I}{\mathcal{I}}
\renewrobustcmd{\P}{\mathcal{P}}   
\newrobustcmd{\W}{\mathcal{W}}
\newrobustcmd{\R}{\mathds{R}}
\newrobustcmd{\Temp}{T}
\newrobustcmd{\dual}[1]{\bar{#1}}
\newrobustcmd{\one}{\mathds{1}}
\newrobustcmd{\ket}[1]{|#1\rangle}
\newrobustcmd{\bra}[1]{\langle#1|}
\newrobustcmd{\brkt}[1]{\langle #1 \rangle}
\newrobustcmd{\braket}[2]{\langle #1 | #2 \rangle}
\newrobustcmd{\Ket}[1]{\bm{|}#1\bm{)}}
\newrobustcmd{\Bra}[1]{\bm{(}#1\bm{|}}
\newrobustcmd{\Braket}[2]{\bm{(}#1\bm{|}#2\bm{)}}
\newrobustcmd{\Brkt}[1]{\bm{(} #1 \bm{)}}
\newrobustcmd{\op}[1]{\hat{#1}}
\DeclareMathOperator{\Tr}{Tr}
\newrobustcmd{\tr}{\underset{\res}{\Tr}}
\newrobustcmd{\tri}{\Tr_\res}
\newrobustcmd{\col}[4]{{\begin{bmatrix}#1 \\ #2 \\ #3 \\ #4 \end{bmatrix}}}
\newrobustcmd{\row}[4]{{\begin{bmatrix}#1 &  #2  & #3  & #4 \end{bmatrix}}}
\newrobustcmd{\suppmat}{\cite{Schulenborg15Suppmat}}
\newrobustcmd{\Eq}[1]{Eq.~(\ref{#1})}
\newrobustcmd{\eq}[1]{(\ref{#1})}
\newrobustcmd{\Fig}[1]{Fig.~\ref{#1}}
\newrobustcmd{\fig}[1]{\ref{#1}}
\newrobustcmd{\Figs}[1]{Figs.~\ref{#1}}
\newrobustcmd{\Sec}[1]{Sec.~\ref{#1}}
\newrobustcmd{\Ref}[1]{Ref.~\cite{#1}}
\newrobustcmd{\Refs}[1]{Refs.~\cite{#1}}
\newrobustcmd{\App}[1]{App.~\ref{#1}}
\newrobustcmd{\fpop}{(-\one)^N}

\newrobustcmd{\zin}{\bar{z}}

\newrobustcmd{\Phif}{\Phi_\mathrm{f}}
\newrobustcmd{\Phil}{\Phi_\mathrm{l}}
\newrobustcmd{\KetUp}{\Ket{\!\uparrow}}
\newrobustcmd{\KetDown}{\Ket{\!\downarrow}}
\newrobustcmd{\BraUp}{\Bra{\uparrow\!}}
\newrobustcmd{\BraDown}{\Bra{\downarrow\!}}
\newrobustcmd{\ketUp}{\ket{\!\uparrow}}
\newrobustcmd{\ketDown}{\ket{\!\downarrow}}
\newrobustcmd{\braUp}{\bra{\uparrow\!}}
\newrobustcmd{\braDown}{\bra{\downarrow\!}}
\newrobustcmd{\sudag}{{\boldsymbol{\dagger}}}
\newrobustcmd{\KetTilC}{\Ket{\tilde{c}}}
\newrobustcmd{\KetTilS}{\Ket{\tilde{s}}}
\newrobustcmd{\BraTilC}{\Bra{\tilde{c}'}}
\newrobustcmd{\BraTilS}{\Bra{\tilde{s}'}}
\newrobustcmd{\gamTilC}{\gamma_{\tilde{c}}}
\newrobustcmd{\gamTilS}{\gamma_{\tilde{s}}}
\newrobustcmd{\gamMixC}{\gamma^{\mix}_{c}}
\newrobustcmd{\gamMixS}{\gamma^{\mix}_{s}}
\newrobustcmd{\DelGamC}{\Delta\gamma_{c}}
\newrobustcmd{\DelGamS}{\Delta\gamma_{s}}
\newrobustcmd{\gamCUp}{\gamma_{c\uparrow}}
\newrobustcmd{\gamSUp}{\gamma_{s\uparrow}}
\newrobustcmd{\gamCDown}{\gamma_{c\downarrow}}
\newrobustcmd{\gamSDown}{\gamma_{s\downarrow}}
\newrobustcmd{\gamCSig}{\gamma_{c\sigma}}
\newrobustcmd{\gamSSig}{\gamma_{s\sigma}}
\newrobustcmd{\BrackEq}[1]{$[$\Eq{#1}$]$}
\newrobustcmd{\ThetaC}{\Theta}
\newrobustcmd{\todo}[1]{\textbf{\textcolor{red}{TODO: #1}}}
\newrobustcmd{\HL}[1]{\textcolor{red}{#1}}

\renewrobustcmd{\todo}[1]{}

\renewrobustcmd{\Fig}[1]{{Fig.~\ref{#1}}}
\renewrobustcmd{\fig}[1]{{\ref{#1}}}
\newrobustcmd{\mix}{}
\newrobustcmd{\jens}[1]{\textcolor{red}{JENS: #1}\textcolor{black}}

\begin{document}

\title{Relaxation of quantum dots in a magnetic field at finite bias
\\ -- 
charge, spin and heat currents}

\titlerunning{Relax. ferm. sys.}

\author{%
  Joren Vanherck\textsuperscript{\textsf{\bfseries 1,2}},
  Jens Schulenborg\textsuperscript{\Ast,\textsf{\bfseries 1}},
  Roman B. Saptsov\textsuperscript{\textsf{\bfseries 3}}, 
    Janine Splettstoesser\textsuperscript{\textsf{\bfseries 1}},
   Maarten R. Wegewijs\textsuperscript{\textsf{\bfseries 4}}}

\authorrunning{Joren Vanherck, Jens Schulenborg, Roman B. Saptsov, Janine Splettstoesser, Maarten R. Wegewijs}

\mail{e-mail
  \textsf{jenssc@chalmers.se}}

\institute{%
  \textsuperscript{1}\,Department of Microtechnology and Nanoscience (MC2), Chalmers University of Technology, SE-41298 G{\"o}teborg, Sweden\\
  \textsuperscript{2}\,Physics Department, Universiteit Antwerpen, B-2020 Antwerpen, Belgium\\
  \textsuperscript{3}\,Institute for Theory of Statistical Physics, RWTH Aachen, 52056 Aachen,  Germany\\
  \textsuperscript{4}\,Peter Gr{\"u}nberg Institut, Forschungszentrum J{\"u}lich, 52425 J{\"u}lich,  Germany \& JARA-FIT}
  
\received{XXXX, revised XXXX, accepted XXXX} 
\published{XXXX} 


\abstract{%
\abstcol{%
We perform a detailed study of the effect of finite bias and magnetic field on the tunneling-induced decay of the state of a quantum dot by applying a recently discovered general duality [PRB  \textbf{93}, 81411 (2016)]. This duality provides deep physical insight into the decay dynamics of electronic open quantum systems with strong Coulomb interaction. It associates the amplitudes of decay eigenmodes of the actual system to the eigenmodes of a so-called dual system with \emph{attractive} interaction. Thereby, it predicts many surprising features in the transient transport and its dependence on experimental control parameters: the attractive interaction of the dual model shows up as sharp features in the amplitudes of measurable time-dependent currents through
}{
the actual repulsive system. In particular, for interacting quantum dots, the time-dependent heat current exhibits a decay mode that dissipates the interaction energy and that is tied to the fermion parity of the system. We show that its decay amplitude has an unexpected gate-voltage dependence that is robust up to sizable bias voltages and then bifurcates, reflecting that the Coulomb blockade is lifted in the dual system. Furthermore, combining our duality relation with the known Iche-duality, we derive new symmetry properties of the decay rates as a function of magnetic field and gate voltage. Finally, we quantify charge- and spin-mode mixing due to the magnetic field using a single mixing parameter.
}}

\maketitle   

\section{Introduction}
The last decades have seen impressive advances in the miniaturization of electronic devices, reaching device sizes on the molecular scale. To develop new functionalities that benefit from this strong size-confinement, the understanding and control of the \emph{single-} and \emph{few-}electron effects governing the nanoscale play a key role. In recent years, low-temperature experiments on nanoscale electronic devices have demonstrated the ability to manipulate single electrons, as reviewed in, e.g., \Ref{Pekola13rev}. To make such setups useful for applications, it is typically required that an external agent operates the system in a time-dependent manner. The speed at which these operations can be performed crucially depends on the various relaxation times of the system, i.e., the characteristic times for the system to make a transition between different quantum states.\par

Thus, the study of these times is of great importance for future developments in nanoelectronics. In particular, it is necessary to understand how they enter measurable quantities -- most importantly time-dependent charge, spin and heat currents -- and how they depend on the device specifics, external control parameters, and non-equilibrium conditions. This knowledge yields a direct input for a better characterization of present, and enhanced design of future nanosystems intended for electronic and spintronic applications. It also includes an improved control over inevitable heating effects due to the time-dependent operation.\par

Experimentally, the time scales of charge emission and absorption on the single-electron level have been studied in various single- and multiple quantum dot setups~\cite{Fujisawa98,Feve07,Roche13}. The emission and reabsorption of single-spin currents has been experimentally implemented, using surface acoustic waves for single-electron control~\cite{Bertrand16}. Recently, the energy-resolved current out of a quantum dot in response to a rapidly modulated confining potential inducing clock-controlled emission of one or two electrons has been measured~\cite{Waldie15,Ubbelohde15}. With new techniques being developed for, e.g., fast thermometry~\cite{Gasparinetti15}, also the time-resolved detection of \emph{heat currents} carried by single electrons is coming within reach.\par

In highly confined electronic systems such as the ones mentioned above, the spacing between energy levels can typically be much larger than the energy scales set by temperature, bias, and driving parameters such as gate voltages and magnetic fields. This confinement on the one hand simplifies matters, as it is often a good approximation to model a generic small quantum device as a few-level quantum dot tunnel coupled to electronic reservoirs. On the other hand, the smaller size also increases the Coulomb interaction strength between electrons occupying the quantum dot. This makes the interaction an equally important ingredient for an experimentally relevant model. Finally, the coupling between quantum system and electronic reservoirs is often small compared to temperature and bias. 
This regime of weak coupling is particularly interesting for the purpose of single- and few-electron control considered here: the weak coupling ensures that as long as electrons are inside the electronic device, they are affected mainly by \emph{local and well tunable} control parameters, and less by the environment.\par

Theoretically analyzing the time evolution due to a change of control parameters of the device is a challenging task. The problem is that even the simple model systems discussed here are nevertheless interacting and open many-body quantum systems. Furthermore, they are generally not in equilibrium with the external electronic reservoirs to which they couple. Within a systematic approach to tackle these problems, we have recently put forward a new nontrivial property of kinetic equations governing the density operator of open fermion systems prepared in an initially nonstationary state. This property takes the form of a \emph{duality relation} that generalizes the hermiticity of time-evolution generators for closed systems to those generating nonunitary open-system evolution~\cite{Schulenborg16}. 
The duality has a broad validity regime, applying to a large class of fermionic quantum systems, which can possibly be strongly coupled to external fermionic reservoirs and exhibit memory effects. Its usefulness consists in technical advantages for the calculation of relaxation properties -- decay modes, amplitudes, and relaxation times -- as well as in the physical intuition that the duality establishes for these properties and their origins. While the general duality relation has a complicated form, it becomes particularly simple and insightful for weakly coupled quantum systems of interest here. For such systems, the kinetic equation simplifies to a Born-Markov master equation~\cite{Breuer}, for which the duality relation in particular sheds light on the physics of the ``fermion-parity'' decay mode. This mode was only recently noted~\cite{Contreras12,Schulenborg14a,Saptsov12a,Saptsov14a}, despite many previous studies of the underlying simple master equation.
By now, it has been shown to be responsible for unexpected features of the time-resolved heat current emitted from a driven interacting quantum dot in contact with a single electronic reservoir~\cite{Schulenborg16}.\par 

In this paper, we first extensively review the duality relation in the Born-Markov limit in \Sec{sec::duality}. In section \ref{sec::quantum_dot}, we then extend the application of this duality relation beyond the scope of \Ref{Schulenborg16} by studying the impact of two voltage-biased leads, and of an externally applied magnetic field. We find also for this quite general case that the relaxation behavior of a quantum dot subject to electron tunneling induced by a sudden parameter change exhibits a remarkable robustness of the fermion-parity mode. Furthermore, we combine the duality relation with the known Iche-duality~\cite{Iche72} to derive a new symmetry of the relaxation mechanism in a magnetic field. We also discuss experimentally relevant protocols for the study of the impact of the magnetic field on the decay behavior and show that this can be characterized by a single mode-mixing parameter. 
Section \ref{sec::outlook} finally discusses further extensions and future directions.

\section{Duality relation for open electronic systems\label{sec::duality}}

\subsection{General framework}
\label{sec::general::mode_amplitude_duality}
To introduce the duality relation in a clear way, we must first review the open-system formulation of the dynamics of electronic devices. The system of interest is a single-electron control device that forms an \emph{open} subsystem of a larger, closed system which also includes electronic leads used to access the subsystem via tunnel coupling. The quantum state of the total closed system evolves unitarily from an initial state $\rho^\tot_0$ as
\begin{align}
\rho^\tot(t)=e^{-i H^\tot t}\rho^\tot_0 e^{+i H^\tot t}\label{eq::general::evolution_tot}
\end{align}
(we set $\hbar = e = k_\mathrm{B} = 1$),
governed by a Hamiltonian
\begin{equation}
 H^\tot = H + H^\res + H^\tun.\label{eq::general::hamiltonian_tot}
\end{equation}
This total Hamiltonian includes an arbitrary electronic Hamiltonian $H$ for the subsystem, an effectively non-interacting lead Hamiltonian 
\begin{equation}
H^\res = \sum_{r,\boldsymbol{k},\sigma,n}\epsilon_{r\sigma n}(\boldsymbol{k})c^\dagger_{r\boldsymbol{k}\sigma n}c_{r\boldsymbol{k}\sigma n}, 
\end{equation}
where $c^\dagger (c)$ create (annihilate) electrons in the energy bands $\epsilon_{r\sigma n}(\boldsymbol{k})$ depending on the orbital index $\boldsymbol{k}$, lead $r$, spin $\sigma=\uparrow,\downarrow$ and band index $n$. Finally, we assume a bilinear lead-system electron tunneling Hamiltonian
\begin{equation}
 H^\tun = \sum_{r,\boldsymbol{k},\sigma,n,l}\tau_{rnl\sigma}(\boldsymbol{k})c^\dagger_{r\boldsymbol{k}\sigma n}d_{l\sigma} + \text{H.c.}\,.\label{eq::general::hamiltonian_tun}
\end{equation}
The latter contains the overlaps $\tau_{rnl\sigma}(\boldsymbol{k})$ between the energy bands in the leads and the single-particle states $l,\sigma$ defining the subsystem. The operators $d^\dagger_{l\sigma}$ and $d_{l\sigma}$ create and annihilate electrons in these states. We allow for \emph{spin-dependent} tunneling, which conserves the component ($\sigma$) along one fixed axis (e.g., an external magnetic field).

\paragraph{Relaxation of electronic open systems}

While Eqs.~\eq{eq::general::evolution_tot}-\eq{eq::general::hamiltonian_tun} in principle completely characterize the unitary time evolution of the highly complex total system, our focus is only on the time-dependent properties of the open subsystem. These are encoded in the reduced density operator, obtained by taking the partial trace over the reservoir degrees of freedom: ${\rho(t)} = \Tr_{\mathrm{lead}}\rho^\tot(t)$. This subsystem quantum state evolves nonunitarily, since the tunnel coupling causes ${\rho(t)}$ to decay to a stationary-state or \emph{zero-mode} operator $z :=\ \lim_{t\rightarrow\infty}{\rho(t)}$. Physically understanding this nonunitary decay in electronic devices in terms of decay modes and their decay rates is the central topic of this article.\par

To obtain ${\rho(t)}$ in practice, one commonly first derives a kinetic or generalized master equation~\cite{Bloch53,Schoeller94,Koenig96b,Timm08,Schoeller09b,Cohen13a,Karlewski14}. Modeling the leads as reservoirs that are initially each in a grand-canonical equilibrium state contained in $\rho^\res_0$, and furthermore assuming no initial system-environment correlations, $\rho^\tot_0 = \rho_0 \cdot \rho^\res_0$, the relaxation dynamics for any time $t > 0$ is described by the quantum kinetic equation $\partial_t{\rho(t)} = -iL{\rho(t)} + \int_0^tdt'\W(t - t'){\rho(t')}$. The Liouvillian $L=[H,\bullet]$, the commutator with the subsystem Hamiltonian $H$, by itself generates unitary dynamics; the kernel $\W(t - t')$ accounts for the effects of coupling between system and leads via tunneling, resulting in nonunitary evolution. 
This coupling in principle also gives rise to time-nonlocal, non-Markovian dynamics~\cite{Breuer,Schoeller09a,Breuer12}, as the environment acts as a memory for the open system. In this work, we focus on \emph{weakly coupled} systems, for which the typical tunneling times set by $H^\tun$ and $H^\res$  are large compared to the typical memory times set by the lead temperatures $T_r$. For such setups, one can expand the coupling kernel $\W$ up to the leading (quadratic) order in the tunneling $H^\tun$, $\W \approx \W_2$, and furthermore neglect non-Markovian effects, leading to the \emph{Born-Markov master equation}
\begin{equation}
 \partial_t{\rho(t)} = W{\rho(t)} \quad,\quad W =  \lim_{\eta\searrow 0} \int_0^\infty dt\W_2(t)e^{-\eta t}.\label{eq::general::born_markov_equation}
\end{equation}
In writing \Eq{eq::general::born_markov_equation}, we restrict our attention\footnote{\Ref{SchulenborgLicentiate} extends the present discussion to include the coherent dynamics neglected here.} to situations for which only occupations of the subsystem energy ($H$) eigenstates matters. For the part of $\rho(t)$ that describes this, we have $L\rho(t) = [H,\rho(t)]=0$, i.e., we can drop $L$.\par

\paragraph{Comparing open and closed system dynamics}

Given an initial condition ${\rho_\In}$, the solution to \Eq{eq::general::born_markov_equation} has the same exponential form as for closed systems, $\rho^\text{closed}(t) = e^{-iLt}\rho_\In^\text{closed}$, but with the coupling kernel $W$ generating the time evolution
\begin{align}
\label{eq::general::reducedMarkovevolution}
 \rho(t)=e^{W t}\rho_\In.
\end{align}
While formally similar, the decisive difference between closed and open systems is that $L$ generates unitary, and $W$ nonunitary dynamics. As illustrated more precisely below, the fact that unitarity dictates $L$ to be hermitian allows for the systematic and intuitive physical description of time evolution in arbitrary closed systems that we are accustomed to. For the nonunitary evolution due to $W$, few such general relations are known. The duality discussed here is, to our knowledge, the most basic relation onto which a similarly \emph{systematic} approach to describing \emph{dissipative} open system dynamics in a non-equilibrium setup can be based.\par

To formulate the duality clearly, we need some more suitable notation. Realizing that the set of operators acting on the Hilbert space of the open system forms a vector space -- the so-called Liouville space -- we denote~\cite{Breuer,Saptsov12a,Schulenborg16} an operator $x$ as a ``ket'' vector $\Ket{x} = x$. Via the Hilbert-Schmidt scalar product, we can let a vector $\Ket{x}$ act on another vector $\Ket{\bullet}$ to yield a number: with $\bullet$ denoting an operator argument, we define a ``bra'' or covector $\Bra{x}\bullet = \Tr[x^\dagger\bullet]$. For superoperators $\A$, such as $W$ or $-iL$, the \emph{Liouville-space hermitian conjugate} with respect to this scalar product is then defined as usual: $\Bra{x}\A^\sudag\Ket{y} := \Bra{y}\A\Ket{x}^*$. In this sense, a rounded bra can also be understood as the Liouville-adjoint of the ket $\Ket{x}$, i.e., $\Bra{x}=\left[ \Ket{x} \right]^\sudag$.
Using this notation, any time evolution kernel can be expanded in terms of bra and ket basis vectors: $\A=\sum_{ij}A_{ij}\Ket{x_i}\Bra{x_j}$.\par

To become more familiar with this formulation, we now first use it for the case of unitary closed-system evolution, $\A=-iL=-i[ H,\bullet]$. For this case,  $L$ is hermitian, $L^\sudag = L$, implying real eigenvalues. If we denote by $\ket{E_i}$ the many-particle energy eigenstates of the system \emph{Hamiltonian} $H$ with energy $E_i$, then the eigenvalues of $L$ are all possible differences $E_i - E_j \in \R$. The corresponding right eigenvectors are $\Ket{E_{ij}} := \ket{E_i}\bra{E_j}$. Importantly, the  hermiticity of $L$ now dictates that the left and right eigenvectors of $L$ to the same eigenvalue $E_i - E_j$ are Liouville-space adjoints. Using this property, we recover the well-known state evolution in the energy eigenbasis of a closed quantum system: with $\Braket{E_{ij}}{\rho_\In}=\bra{E_i} \rho_\In \ket{E_j}$, one finds
\begin{equation}
 \Ket{\rho(t)} = e^{-iLt}\Ket{\rho_\In}
=\sum_{ij}\Braket{E_{ij}}{\rho_\In}\cdot e^{-i(E_i - E_j)t}\Ket{E_{ij}}.\label{eq::general::energy_expansion}
\end{equation}
If the state was prepared in a mixture (no coherent superpositions), all oscillating terms $i \neq j$ are zero, and the remaining $i=j$ terms do not evolve. Most important in our context is, however, that since left and right eigenvectors are Liouville-adjoints, the contribution to the evolution $\Ket{\rho(t)}$ of each mode of $L$  is weighted by the overlap $\Braket{E_{ij}}{\rho_\In}$ of the \emph{mode} with the initial state. The intuition is thus: to ``excite'' a certain mode of evolution, prepare the initial state $\rho_\In$ so as to resemble the targeted mode of the \emph{actual} system, similar to, e.g., plucking a string.\par

For the nonhermitian kernel $W$ generating nonunitary dynamics, the above intuitive physical interpretation of the eigenmode expansion partially breaks down. Formally, $\Ket{\rho(t)}$ can still be expanded in the right eigenbasis, i.e., $W\Ket{x}=-\gamma_{x} \Ket{x}$, as follows:
\begin{align}
\label{eq::general::mode_expansion}
 \Ket{\rho(t)} = e^{Wt}\Ket{\rho_\In}
= \sum_{x} \Braket{x'}{\rho_\In}\cdot e^{-\gamma_x t}\Ket{x}.
\end{align}
Each \emph{mode} $\Ket{x}$ decays with a nonnegative rate $\gamma_x$, i.e., it is a possible ``motion of the system'' governed by a single time scale. To the same eigenvalue also belongs a left eigenvector, $\Bra{x'}W=-\gamma_{x} \Bra{x'}$. The \emph{amplitude} of mode $\Ket{x}$ in the full time evolution is determined by the overlap of this corresponding left eigenvector $\Bra{x'}$ with the initial state $\Ket{\rho_\In}$. Importantly, the symbol $x'$ in \Eq{eq::general::mode_expansion} denotes a \emph{different} operator than $x$: the labels $x$, $x'$ only indicate that they belong to the \emph{same eigenvalue} $-\gamma_{x}$. The crucial difference to closed systems
\footnote{Similar to the closed-system case, the modes and amplitude covectors are orthonormal, i.e., $\Braket{x'}{y}=\delta_{x,y}$, and decompose the Liouville space identity as $\I = \sum_x\Ket{x}\Bra{x'}$.} 
is that modes and amplitude covectors are not anymore related by Liouville-space hermitian conjugation, $\Bra{x'} \neq \left[\, \Ket{x} \, \right]^\sudag$. To excite mode $\Ket{x}$, one should thus \emph{not} compare the initial state with this targeted mode, but with ``something else'' denoted by $x'$. In other words, there seems to be no obvious \emph{physical} and universally applicable relation between the modes and the amplitude covectors with which these modes enter the reduced dynamics, given an initial state. This is the key obstacle on the way to developing any systematic physical intuition about the relaxation properties of open systems.

\begin{figure}[t]
\centering
\includegraphics[width=0.9\columnwidth]{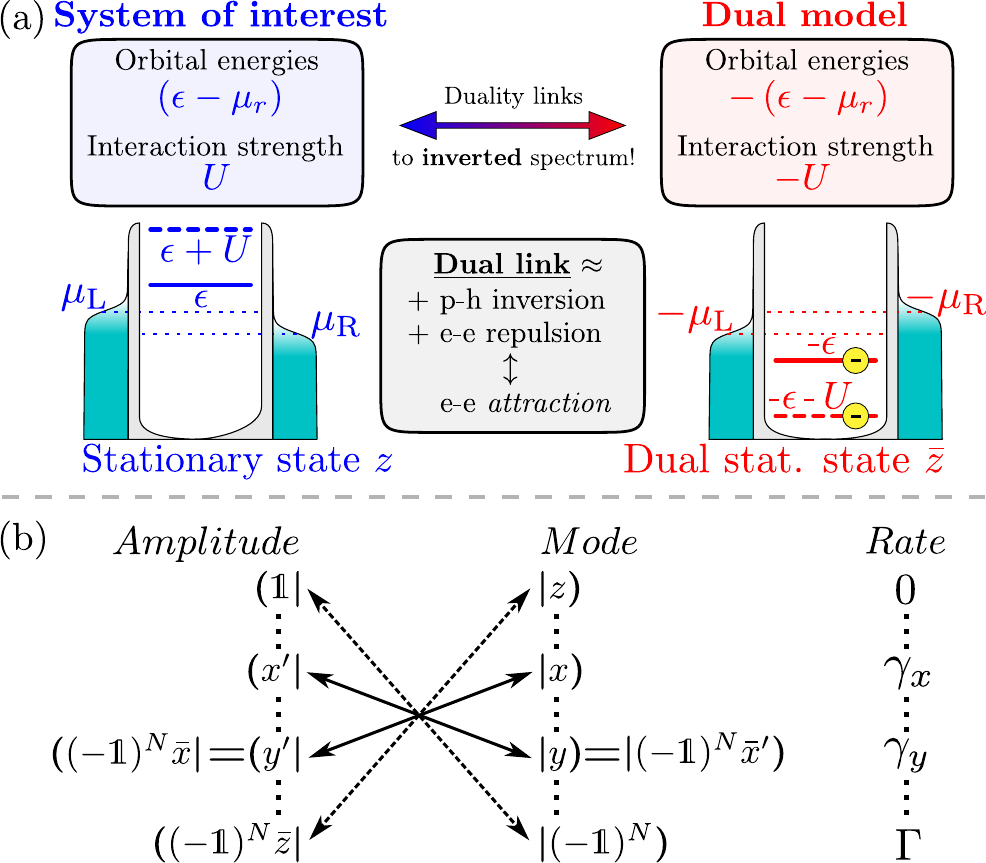}
\caption{Mode-amplitude duality of relaxation dynamics: (a)~The duality \eq{eq::general::duality_relation} links the actual model of interest to a dual model with inverted signs for all transition energies. (b)~\Eq{eq::general::duality_relation} \emph{cross}links the decay modes and the amplitude covectors and their corresponding decay rates as indicated.}
\label{fig::general::duality}
\end{figure}

\paragraph{Mode-amplitude duality}
The surprising result of \Ref{Schulenborg16} is that such a physical connection between modes and amplitude covectors indeed turns out to exist for a large class of fermionic systems. Namely, this applies to systems that (i)~evolve in time according to a Hamiltonian of the form \eq{eq::general::hamiltonian_tot}-\eq{eq::general::hamiltonian_tun}, (ii)~contain noninteracting electronic leads $r$ that are initially uncorrelated with the subsystem, and each in equilibrium (temperatures $T_r$, electrochemical potentials $\mu_r$), and (iii)~have tunneling frequencies
\begin{equation}
\Gamma_{rl\sigma}(\omega) = 2\pi\sum_{\boldsymbol{k},n}\delta(\epsilon_{rn\sigma}(\boldsymbol{k}) - \omega)|\tau_{rnl\sigma}(\boldsymbol{k})|^2 \rightarrow \Gamma_{rl\sigma}\label{eq::general::bare_couplings}
\end{equation}
that are energy$(\omega)$-independent (wideband limit). For these systems, we have derived that the Born-Markov kernel in \Eq{eq::general::born_markov_equation} obeys
\begin{gather}
W^{\sudag} = -\Gamma - \P \dual{W}\P\quad,\quad \dual{W} = W(-H,\{-\mu_r\}) \label{eq::general::duality_relation}.
\end{gather}
This important result in a way generalizes the hermiticity of evolution generators of closed systems, $(iL)^\sudag = -iL$, to open systems. Namely, instead of simply equating $W^\sudag$ to $-W$, the right hand side of \Eq{eq::general::duality_relation} introduces further elements: (1)~a shift given by the lumped sum of all bare couplings, $\Gamma = \sum_{rl\sigma}\Gamma_{rl\sigma} > 0$, (2)~a superoperator $\P\bullet$ which multiplies a subsystem operator $\bullet$ from the left with the fermion-parity operator, $\P=\fpop \bullet = e^{i\pi N}\bullet$ with the subsystem number operator $N = \sum_{l\sigma}d^\dagger_{l\sigma}d_{l\sigma}$, and (3)~a dual kernel $\dual{W}$. This dual kernel is actually a physical kernel for a \emph{dual system}, obtained from a given kernel $W$ for the actual system by performing a simple parameter transform: 
invert the sign of all energies, $H \to -H$, and of all electrochemical potentials, $\mu_r \to -\mu_r$, for all leads $r$.This duality transformation is sketched in \Fig{fig::general::duality}(a). In the following, we will indicate all quantities relating to this dual system -- obtained by the same substitution -- by an overbar, e.g., $H \to \dual{H}:=-H$.\par

The duality \eq{eq::general::duality_relation} establishes a practical and physically meaningful relation between modes and amplitude covectors of $W$. Given a mode $\Ket{x}$  with rate $\gamma_x$, one finds that $\Bra{y'} = \Bra{\fpop\dual{x}}$ is an amplitude covector for a decay rate $\gamma_{y}$:
\begin{gather}
 W\Ket{x} = -\gamma_x\Ket{x} \Rightarrow \Bra{y'}W = -\gamma_y\Bra{y'}\notag\\
 \Bra{y'} = \Bra{\dual{x}}\P = \Bra{\fpop\dual{x}} \quad,\quad \gamma_y = \Gamma - \dual{\gamma}_x.\label{eq::general::modeToAmplitude}
\end{gather}
Likewise, the existence of an amplitude covector $\Bra{x'}$ dictates that $\Ket{y} = \Ket{\fpop\dual{x}'}$ is a mode. The duality \eq{eq::general::duality_relation} thus nontrivially \emph{crosslinks} modes and amplitude covectors to generally \emph{different} eigenvalues, via the fermion-parity $\fpop$ and the dual model specified by $-H$ and $\{-\mu_r\}$. This is illustrated in \Fig{fig::general::duality}(b). The remainder of this article is dedicated to the implications of this crosslink, providing new, experimentally relevant  ideas pertaining to single-electron control, but also strongly simplifying calculations of the time-dependent reduced density operator $\Ket{\rho(t)}$ and observable currents.

\subsection{Fermion-parity mode}
\label{sec::general::fermion_parity_mode}
Even without referring to a specific model, the duality relation provides two main insights into relaxation in open system dynamics. First, one knows that as a mere consequence of probability conservation, any coupling kernel $W$ has the zero left eigenvector given by the unit trace, $\Bra{\one}W\bullet = \Tr\left[W\bullet\right] = 0$. Equation \eq{eq::general::duality_relation} then immediately leads to the existence of what we call the fermion-parity mode~\cite{Contreras12,Saptsov12a,Saptsov14a,Schulenborg14a,Schulenborg16}:
\begin{equation}
 W\Ket{\fpop} = -\Gamma\Ket{\fpop}.\label{eq::general::fermion_parity_mode}
\end{equation}
The corresponding fermion-parity rate equals the sum of bare couplings $\Gamma$, and is hence robust to \emph{any} parameter change except for those affecting this sum. This is fundamentally interesting, since an experiment aimed at exposing this robustness, similar to the theoretical proposals given in \Refs{Contreras12,Schulenborg14a,Schulenborg16}, could test the validity of our duality \eq{eq::general::duality_relation}. Practically, this finding is important for any application that critically relies on the sensitivity of decay times to external influences.\par

Another interesting property of the parity rate $\Gamma$ which follows from the generalized hermiticity relation \eq{eq::general::duality_relation} is that it is the \emph{largest} among all the rates $\gamma_x$ entering the decay mode expansion \eq{eq::general::mode_expansion} of the reduced density operator~\cite{Schulenborg16,SchulenborgLicentiate}. This represents an interesting contrast to systems in contact with a bosonic bath instead of a fermionic one, such as for the dissipative two-state system weakly coupling to photons~\cite{Legget87}. 
For the latter setup, spontaneous emission sets a \emph{lower} bound on the relaxation rate that only depends on the bare coupling strength, while increasing temperatures can lead to an arbitrarily high photon number in the bath states which can also arbitrarily increase the decay rate via absorption and stimulated emission. The crucial difference to the fermionic setups considered here is the Pauli principle: it limits both emission and absorption of electrons by allowing only a single particle to tunnel to or from each energetically accessible single-particle state. This can cause a relaxation rate to approach zero, as shown in \Sec{sec::quantum_dot::finite_bias}, and instead sets an \emph{upper} bound that is naturally given by the sum of all tunneling frequencies $\Gamma$.\par

Finally, the fact that this mode is given by the fermion parity operator $\fpop$ has an important implication for few-electron decay dynamics: given an open system with $M$ single-particle states (spin-orbitals), an expectation value $\langle x\rangle(t) = \Braket{x}{\rho(t)}$  of some operator $x$ can only pick up the contribution of $\Ket{\rho(t)}$ that decays with the parity rate $\Gamma$ if $x$ is an $M$-particle observable\footnote{The overlap $\Braket{x}{\fpop}$ can only be non-zero if $x$ contains a product of all $M$ occupation-number operators.}. The parity mode therefore becomes relevant for the transient decay dynamics of exactly the type of systems that we are interested in -- nanostructures emitting or transmitting more than one electron. Relevant examples are, e.g., electron pumps similar to those presented in~\Refs{Fletcher13,Waldie15,Ubbelohde15}.
For strong confinement, only $M=2$ spin-orbitals matter, and the parity operator is essentially the two-particle interaction [cf. \BrackEq{tildeepsH}]. Because one can ``store'' energy in this interaction, this mode is of primary importance for describing heat currents in quantum dots~\cite{Schulenborg16}.\par

\subsection{The dual model}
\label{sec::general::dual_model}
We now turn to the second main insight offered  by \Eq{eq::general::duality_relation}. From the known fact that the kernel $W$ has at least one left zero eigenvector $\Bra{\one}$, it follows that there is also at least one zero right eigenvector $W\Ket{z} = 0$. This is the \emph{stationary state} or zero-mode of the dynamics. As sketched in \Fig{fig::general::duality}(b), the existence of $\Ket{z}$ implies by the duality \eq{eq::general::modeToAmplitude} that the amplitude with which the parity mode $\Ket{\fpop}$ enters the reduced density operator $\Ket{\rho(t)}$ is determined by the covector $\Bra{\dual{z}\fpop}$:
\begin{equation}
 \Bra{\dual{z}\fpop}W = -\Gamma\Bra{\dual{z}\fpop}.\label{eq::general::dual_state}
\end{equation}
Here, $\dual{z}$ denotes the stationary state operator of the \emph{dual system}: it is obtained from the stationary state $z$ by inverting the signs of all local energies, $H \rightarrow -H$, and of all lead potentials, $\mu_r \rightarrow -\mu_r$. This energy inversion can be physically interpreted as a combination of a particle-hole transform and, interestingly, a sign inversion of all many-body interactions. This exemplifies how the duality \eq{eq::general::duality_relation} generally relates modes and amplitude covectors of a system with repulsive Coulomb repulsion to an effective model governed by an \emph{attractive} electron-electron interaction.\par    

The fact that the operator $\dual{z}$ is the stationary state of a physical system has a crucial advantage: it implies that the parameter-dependence of the amplitude with which the parity mode enters $\Ket{\rho(t)}$ can be understood from physical principles. For example, a consequence of the electron-electron attraction in the dual model is that it leads to pairing as long as the interaction strength is the dominant energy scale~\cite{Anderson75,Taraphder91}: the dual occupation $\Braket{N}{\dual{z}}$ and parity $\Braket{\fpop}{\dual{z}}$ then tend to be even. Such insights will be illustrated in detail in the following.

\section{Relaxation of a single-level quantum dot}
\label{sec::quantum_dot}
\begin{figure}[t]
\includegraphics[width=1.0\columnwidth]{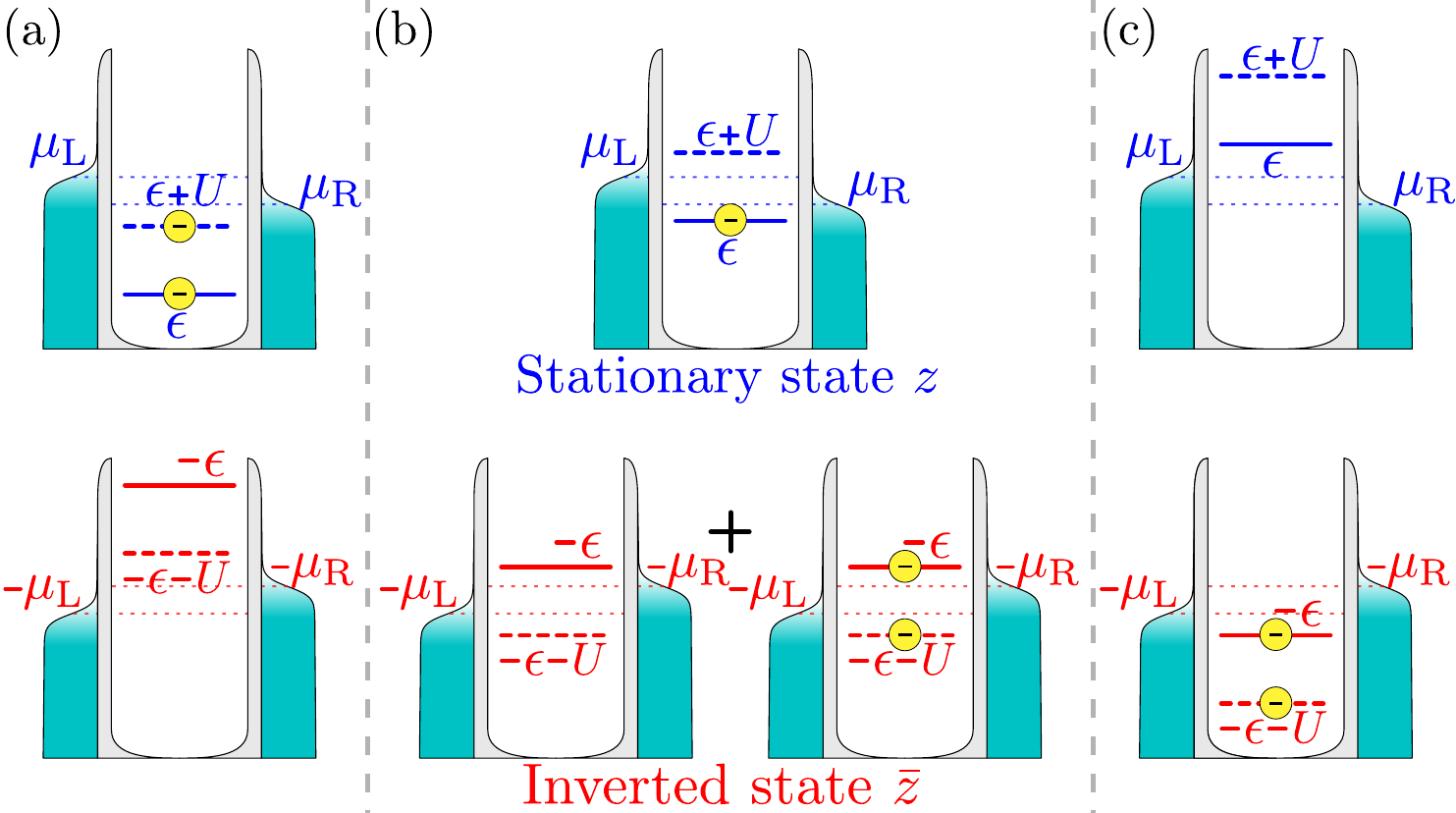}
\caption{Relaxation of a single-level quantum dot. The energy level is denoted by $\epsilon$, the interaction energy by $U$. The dot is tunnel coupled to two leads with electrochemical potentials $\mu_\mathrm{L} \neq \mu_\mathrm{R}$. The occupations in the stationary mixed state $\Ket{z}$ are illustrated in the upper panels for (a)~$\epsilon+U<\mu$, (b)~$\epsilon+U/2=\mu$ and (c)~$\epsilon>\mu$. For each case, the lower panel shows the corresponding situation in the dual model, in which all energies are inverted.} \label{fig::quantum_dot::model_dualmodel}
\end{figure}

The essence of a tunable nanosystem with strong Coulomb interaction is captured by a single-level quantum dot tunnel-coupled to two leads, sketched in the upper panel in \Fig{fig::quantum_dot::model_dualmodel}(a). A conceptually simple way to gain physical insight into the time-dependent decay of such a system is to apply a fast switch operation, and then to infer the decay of the resulting nonstationary state by monitoring the transport currents related to charge, spin and energy in time. In addition to the previous analysis of the transient currents given in \Ref{Schulenborg16}, we here also account for two leads in order to discuss the effect of a voltage bias on the system's transient behavior after the switch. Furthermore, we include the effect of a magnetic field.
After setting up a model, we approach our task in two steps: we first exploit the duality relation \eq{eq::general::duality_relation} to efficiently derive analytical expressions for the quantum state $\Ket{\rho(t)}$. The form of this result is more transparent than that obtained by any straightforward computation. This allows us to discuss in a second step how the time-dependent decay depends on control parameters using a new set of physical arguments based on the dual picture.

\subsection{Model}
\label{sec::quantum_dot::model}

The Hamiltonian $H^\tot$ of the total system is of the general form \eq{eq::general::hamiltonian_tot}-\eq{eq::general::hamiltonian_tun}. The open system consists of a single orbital with spin; its dynamics is described by the (Anderson) model Hamiltonian
\begin{equation}
 H = \epsilon N + B S + U N_\uparrow N_\downarrow.\label{eq::quantum_dot::hamiltonian}
\end{equation}
The externally tunable level position $\epsilon$ and magnetic field $B$ determine the single-particle energies $\epsilon_\sigma = \epsilon + \sigma B/2$ with $\sigma = \uparrow,\downarrow = +,-$ via the total occupation $N = N_\uparrow + N_\downarrow$ and the Zeeman term $B S$ with the operator $S = \frac{1}{2}(N_\uparrow - N_\downarrow)$ for the spin-component along the field. The Coulomb interaction leads to an additional charging energy $U$ in case of double occupation. The operators $N_\sigma = d_\sigma^\dagger d_\sigma$ contain creation $(d^\dagger_\sigma)$ and annihilation $(d_\sigma)$ operators of dot electrons with spin $\sigma$.\par

The two noninteracting electronic leads to which the dot couples are denoted by $r=\mathrm{L,R}$. We define the mean electrochemical potential $\mu := (\mu_\mathrm{R} + \mu_\mathrm{L})/2$ and the bias $V = \mu_\mathrm{L} - \mu_\mathrm{R}$. Since we are interested in the effect of a voltage bias, we assume equal temperatures $T_\mathrm{L}=T_\mathrm{R}=T$ for simplicity. The frequencies for tunneling processes between the dot and the leads are governed by the barrier transparencies $\Gamma_{r\sigma}$, defined in \Eq{eq::general::bare_couplings}. They are assumed to be energy-independent (wideband limit). For later convenience, we also introduce partial sums $\Gamma_r = \sum_\sigma\Gamma_{r\sigma}$, $\Gamma_\sigma = \sum_r\Gamma_{r\sigma}$, and the important lumped sum of all couplings $\Gamma = \sum_r\Gamma_r = \sum_\sigma\Gamma_\sigma$ featuring in the duality \eq{eq::general::duality_relation}.\par

\subsection{Relaxation properties and duality}
\label{sec::quantum_dot::density_operator}
We now determine all relaxation properties of this specific quantum-dot setup in response to a switch.

\paragraph{Time-dependent state}

We consider the weak coupling regime, $\Gamma_{r\sigma} \ll T$, for which the time evolution of the dot state $\Ket{\rho(t)}$ is governed by the Born-Markov master equation \eq{eq::general::born_markov_equation}. The explicit matrix for the kernel $W$ for the above model is written in \App{App::W_Expectations}. The density operator $\rho_\In$ prior to the \emph{instant} change of a system parameter enters the master equation as the initial condition: $\rho(t=0) = \rho_\In$. [Its form needs to be specified only later on, cf. \Eq{eq::quantum_dot::particle_current_leadr}.] In practice, this is realized by performing the switch operation  fast on the scale of the typical tunneling times $1/\Gamma$, giving the dot no time to adjust its state to the new parameters. Relative to the kernel $W$ -- set by the parameters \emph{after} the switch --  the initial state $\rho_\In$ is thus nonstationary, resulting in decay of the form \eq{eq::general::mode_expansion}:
\begin{align}
 \Ket{\rho(t)} =& \Ket{z} + \Braket{p'}{\rho_\In}\cdot e^{-\gamma_p t}\Ket{p}\notag\\
 &+ \Braket{\tilde{c}'}{\rho_\In}\cdot e^{-\gamTilC t}\KetTilC + \Braket{\tilde{s}'}{\rho_\In}\cdot e^{-\gamTilS t}\KetTilS.\label{eq::quantum_dot::reduced_density_operator}
\end{align}
There are four modes since we solve the master equation \eq{eq::general::born_markov_equation} in the subspace spanned by pure-state operators for zero occupation $\Ket{0} := \ket{0}\bra{0}$, spin up/down $\Ket{\sigma} := \ket{\sigma}\bra{\sigma}$ with $\sigma := \uparrow,\downarrow$, and double occupation $\Ket{2} := \ket{2}\bra{2}$. The unique~\cite{SchulenborgLicentiate} stationary state or zero mode of the kernel $W$ is denoted by $\Ket{z} = \lim_{t\rightarrow\infty}\Ket{\rho(t)}$; the vectors $\Ket{p},\KetTilC,\KetTilS$ and $\Bra{p'},\BraTilC,\BraTilS$ are the decay eigenmodes and corresponding amplitude covectors to the decay rates $\gamma_p,\gamTilC,\gamTilS$.\par

In contrast to conventional ``brute force'' diagonalization, we here consider \Eq{eq::quantum_dot::reduced_density_operator} just as an \emph{ansatz} for the form of the time-dependent dot state $\Ket{\rho(t)}$, and see how far we can \emph{construct it by the duality} \eq{eq::general::duality_relation}. Here, this duality takes the concrete form
\begin{equation}
W^\sudag(\epsilon,B,U,\mu_r)=-\Gamma-\P W(-\epsilon,-B,-U,-\mu_r) \P.\label{eq::quantum_dot::duality_relation}
\end{equation}
As discussed after \Eq{eq::general::fermion_parity_mode}, this implies that the parity mode must appear in \Eq{eq::quantum_dot::reduced_density_operator}, denoted by $\Ket{p} := \Ket{\fpop}$ with rate $\gamma_p := \Gamma = \Gamma_\mathrm{L} + \Gamma_\mathrm{R}$. To construct the corresponding  covector $\Bra{p'} = \Bra{\zin\fpop}$ for its amplitude \BrackEq{eq::general::dual_state}, we first need to determine the \emph{non-equilibrium} stationary state $\Ket{z}$ and then obtain the dual stationary state $\Ket{\zin}$ by inverting all energies. Using that $\Ket{\one}, \Ket{N}, \Ket{S}, \Ket{\fpop}$ form an orthogonal basis for the considered Liouville subspace, one can formally expand the mode vector as
\begin{align}
 \Ket{z} &= \left[\frac{1 + p_z}{4} - \frac{N_z-1}{2}\right]\!\Ket{0} + \left[\frac{1 + p_z}{4} + \frac{N_z - 1}{2}\right]\!\Ket{2}\notag\\
 &+\left[\frac{1 - p_z}{4} + S_z\right]\KetUp + \left[\frac{1 - p_z}{4} - S_z\right]\KetDown.\label{eq::quantum_dot::stationary_state}
\end{align}
This parameterizes the stationary state $\Ket{z}$ in terms of its average particle number $N_z = \Braket{N}{z}$, spin $S_z = \Braket{S}{z}$, and fermion-parity $p_z = \Braket{\fpop}{z}$. (Contrary to usual notation, the operator of the spin-component along the magnetic field axis is labeled as $S$ and not $S_z$! In our case, ``$S_z$'' instead denotes the stationary expectation of $S$.) One can explicitly determine these averages -- and thus the stationary state $z$ -- in terms of Fermi functions  by solving $W\Ket{z} = 0$ and $\Braket{\one}{z} = 1$, see \App{App::W_Expectations}. However, much of our discussion does not rely on these expressions: it suffices to understand the parameter dependence of these three quantities which mostly follows from simple physical considerations.\par

The required dual state $\Ket{\zin}$ can be expanded in the same way, replacing in \Eq{eq::quantum_dot::stationary_state} all quantities with their dual counterparts $\dual{N}_z,\dual{S}_z$ and $\dual{p}_z$.\footnote{Their explicit expressions are obtained by inverting all energy signs, which  boils down to a simple rule of thumb: replace every Fermi function $f(E)=(e^{E}+1)^{-1}$ appearing in the expressions for the stationary averages by $f(-E)=1 - f(E)$ to obtain the corresponding dual quantity.} Using the operator-completeness relation, $\Bra{0} + \BraUp + \BraDown + \Bra{2} = \Bra{\one}$, and $\Braket{p'}{z} = 0$, we find the parametrization
\begin{align}
 \Bra{p'} &= \Bra{\zin\fpop}\notag\\
 &= \frac{2-\dual{N}_z}{2}\Bra{\Delta 0} + \frac{\dual{N}_z}{2}\Bra{\Delta 2} - 2\dual{S}_z\Bra{\Delta S},\label{eq::quantum_dot::parity_covector}
\end{align}
in terms of the deviations $\Bra{\Delta X} := \Bra{X} - \Braket{X}{z}\cdot\Bra{\one}$ relative to stationary expectation values.\par

To determine the remaining two covectors $\BraTilC,\BraTilS$ in \Eq{eq::quantum_dot::reduced_density_operator} we use that they must be orthogonal to the stationary state $\Ket{z}$ and the parity mode $\Ket{\fpop}$ that we have already found. This suggests to construct the amplitude covectors from the two linearly independent single-particle observable operators, charge $\Bra{N}$ and spin $\Bra{S}$, as these fulfill $\Braket{N}{\fpop} = \Braket{S}{\fpop} = 0$ already by the argument given at the end of \Sec{sec::general::fermion_parity_mode}. Enforcing also orthogonality to $\Ket{z}$, the exact amplitude covectors $\BraTilC,\BraTilS$ are then obtained explicitly as linear combinations of
\begin{equation}
 \Bra{c'} = \Bra{N} - N_z\Bra{\one} \quad,\quad \Bra{s'} = 2 \Big( \Bra{S} - S_z\Bra{\one} \Big).\label{eq::quantum_dot::charge_and_spin_amplitude}
\end{equation}
Given these, the duality \eq{eq::quantum_dot::duality_relation} allows to conveniently \emph{construct} the corresponding mode vectors $\KetTilC,\KetTilS$ by a linear combination of the operators
\begin{align}
\Ket{c} &:= \frac{1}{2}\left[\Bra{\dual{c}'\fpop}\right]^\sudag = \frac{1}{2} \fpop \Big[ \Ket{N} - \dual{N}_z \Ket{\one}\Big]\notag\\
\Ket{s} & := -\frac{1}{2}\left[\Bra{\dual{s}'\fpop}\right]^\sudag = \Ket{S} + \dual{S}_z \Ket{\fpop}\label{eq::quantum_dot::charge_and_spin_mode},
\end{align}
with the constant prefactors $\pm1/2$ introduced to fix the normalization $\Braket{c'}{c} = \Braket{s'}{s} = 1$. The bras \eq{eq::quantum_dot::charge_and_spin_amplitude} and kets \eq{eq::quantum_dot::charge_and_spin_mode} are the eigenvectors at zero magnetic field related to charge ($c$) and spin ($s$) which were considered in \Ref{Schulenborg16}. In the present more general case, there is thus a mixing of charge and spin decay modes which we will discuss further in \Sec{sec::quantum_dot::magnetic_field}. This is reflected by the exact relaxation rates,
\begin{align} 
\gamma_{\tilde{c}/\tilde{s}} = (1 - C)\cdot\gamma^{\mix}_{c/s} + C \cdot\gamma^{\mix}_{s/c},\label{eq::quantum_dot::relaxation_rates}\\
 C := \tfrac{1}{2} \left( 1 - \sqrt{1+\tfrac{4\DelGamC\DelGamS}{(\gamMixC - \gamMixS)^2} } \, \right).\label{eq::quantum_dot::coupling_constant}
\end{align}
The rate mixing can be expressed through a single constant $C \in (0,\frac{1}{2})$ which depends on sums and differences
\begin{gather}
\gamMixC = \gamCUp + \gamCDown, \quad \gamMixS = \gamSUp + \gamSDown \notag\\
\DelGamC = \gamCUp - \gamCDown, \quad \DelGamS = \gamSUp - \gamSDown 
\end{gather}
of charge and spin rates, cf. \Refs{Splettstoesser10,Contreras12}:
\begin{equation}
\gamCSig = \frac{\Gamma_\sigma}{2}(f^+_{\epsilon\sigma} + f^-_{U\sigma}),\quad \gamSSig = \frac{\Gamma_\sigma}{2}(f^-_{\epsilon\sigma} + f^+_{U\sigma}).\label{eq::quantum_dot::spin_resolved_rates}
\end{equation}
Here $f^\pm_{\epsilon \sigma}:=f^\pm_\sigma(\epsilon_\sigma)$ and $f^\pm_{U \sigma}:=f^\pm_\sigma(\epsilon_\sigma+U)$ are the spin-resolved electron ($+$) and hole ($-$) occupation functions, which are reservoir-weighted sums $f^\pm_\sigma(E) = \sum_{r=\mathrm{L,R}} ({\Gamma_{r\sigma}}/{\Gamma_\sigma}) f^\pm_r(E)$ of Fermi functions $f^\pm_r(E) = f( \pm(E - \mu_r)/T)$. Remarkably, these explicit expressions satisfy the relation
\begin{equation}
\gamma_{\tilde{c}/\tilde{s}}(\epsilon,B,U,\mu_r) = \Gamma - \gamma_{\tilde{c}/\tilde{s}}(-\epsilon,-B,-U,-\mu_r).\label{eq::self-dual}
\end{equation}
This states that the corresponding decay rates are \emph{self-dual} with respect to \Eq{eq::general::modeToAmplitude}. In accordance, the duality relates modes and amplitudes as $\Bra{\tilde{c}'} \leftrightarrow \Ket{\tilde{c}}$ and $\Bra{\tilde{s}'} \leftrightarrow \Ket{\tilde{s}}$.\par

By exploiting the duality, we have completely determined the nonstationary time-dependent state \eq{eq::quantum_dot::reduced_density_operator} with only minimal input, parameterizing it through the stationary-state values of a few physical observables, $N_z$, $p_z$, 	$S_z$ and $\dual{N}_z$, $\dual{p}_z$, $\dual{S}_z$, through the rates $\gamma_{c\sigma}, \gamma_{s\sigma}$, and through a single rate-mixing parameter $C$.

\paragraph{Time-dependent currents}

It is now straightforward to describe how the quantum dot emits and absorbs charge, spin and energy in time. For the Born-Markov limit, it was derived in~\Ref{Saptsov12a,SchulenborgLicentiate} for spin-conserving tunneling that the time-dependent current out of the system into lead $r$ is given by
\begin{equation}
 I^r_x(t) = -\Bra{x}W_r\Ket{\rho(t)},\label{eq::quantum_dot::time_dependent_currents}
\end{equation}
taking $x = N$ for the charge current $I^r_N(t)$, $x = S$ for the spin current $I^r_S(t)$, and $x=H$, the Hamiltonian, for the energy current $I^r_E(t)$. The kernel $W_r$ that appears here is the part of the full kernel $W = \sum_rW_r$ that contains only the rates for tunneling to lead $r$. Its eigenvalues and eigenvectors can thus be expressed in the same way as for $W$, replacing expressions such as $N_z=\Braket{N}{z}$ by reservoir-specific expressions $N_{z_r}:=\Braket{N}{z_r}$ with a \emph{sub}script $r$. The stationary state $z_r$ and averages over it are computed as before, assuming, however, that the system is only coupled to lead $r$, i.e., $f^{\pm}(x) \to f^{\pm}_r(x)$. Note that the current $I_x^r$ out of lead $r$ into the quantum dot is the result of coupling to \emph{all} leads [\Eq{eq::quantum_dot::time_dependent_currents} depends on $\Ket{\rho(t)}$], which is indicated by using a \emph{super}script $r$.

\subsection{Effect of finite bias}\label{sec::quantum_dot::finite_bias}
We now analyze in detail the effect of a finite bias, $V=\mu_\text{L}-\mu_\text{R}\neq 0$, on the relaxation properties and the time-dependent currents that probe them at zero magnetic field, $B=0$.

\paragraph{Relaxation rates}

\begin{figure}[t]
\includegraphics[width=1.0\columnwidth]{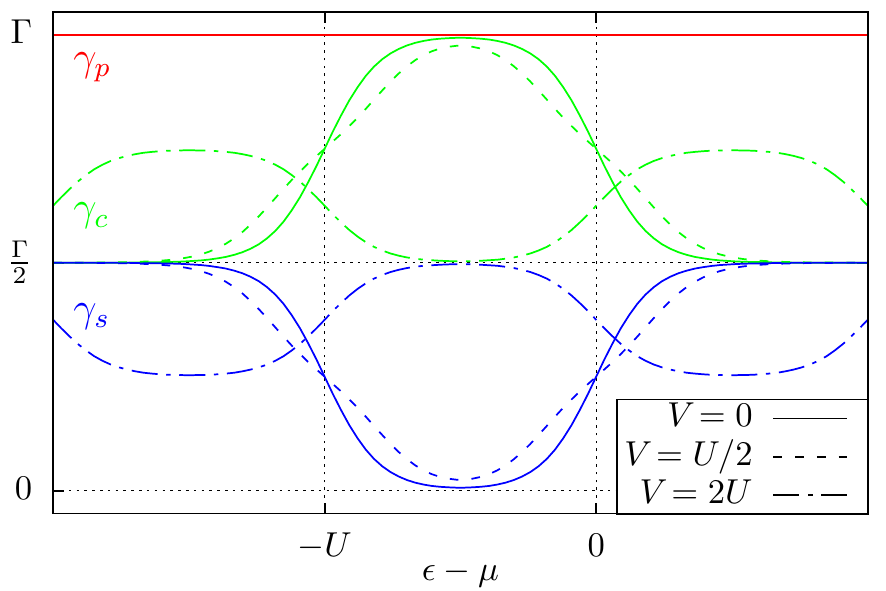}
\caption{Relaxation rates for the parity ($\gamma_p$, red), charge ($\gamma_c$, green) and spin ($\gamma_s$, blue) respectively as a function of the relative dot energy level $\epsilon-\mu$ for three bias values. Parameters: $U=10T$ and $\Gamma_{r\sigma}=\Gamma/4$ for $r=$L,R and $\sigma=\uparrow,\downarrow$.} \label{fig::Decay_Rates}
\end{figure}

At zero field, $B=0$, the relaxation rates are given by the fermion-parity rate $\gamma_p=\Gamma$, and the charge and spin relaxation rates \eq{eq::quantum_dot::relaxation_rates}, $\gamma_{c/s}=\sum_{\sigma=\uparrow \downarrow}\gamma_{c\sigma/s\sigma}$, since the eigenvectors \eq{eq::quantum_dot::charge_and_spin_amplitude}-\eq{eq::quantum_dot::charge_and_spin_mode} are not mixed. In  \Fig{fig::Decay_Rates} we compare the dependence of these rates on the final level position $\epsilon$  for different bias voltages assuming symmetric couplings $\Gamma_{\text{L}\uparrow}=\Gamma_{\text{L}\downarrow}=\Gamma_{\text{R}\uparrow}=\Gamma_{\text{R}\downarrow}=\Gamma/4$. We focus on understanding the values of the plateaus in these plots, ignoring the temperature-smearing at their transitions. When analyzing this $\epsilon$-dependence we highlight two aspects: (i) the magnitude of a certain decay rate depends on the number of accessible decay channels and
(ii) different leads contribute  independently, two spin channels each.\par

The overall scale in \Fig{fig::Decay_Rates} is set by the maximal rate $\gamma_p$. It is robust against the application of any bias since it is independent of $\epsilon, U, \mu_r, T$. This is caused by the fact that  all four decay channels contribute with equal, constant weights causing all energy dependence to cancel out in $\gamma_p$.\par

In contrast, the charge rate $\gamma_c=\sum_{r\sigma}\Gamma_{r\sigma} [ f^+_r(\epsilon)+f^-_r(\epsilon+U)]/2$ strongly depends on the energy level position but also on the bias. For $V=0$, it equals $\Gamma/2$ for $\epsilon$ values for which the dot tends to be empty or doubly occupied. In these regimes, the occupations $N_\uparrow$ and $N_\downarrow$ evolve independently on the time scales $\Gamma_\uparrow = \Gamma/2$ and $\Gamma_\downarrow = \Gamma/2$. Thus, also the sum $N = N_\uparrow + N_\downarrow$ decays on a time scale $\gamma_c = \Gamma/2$. In the singly occupied regime, Coulomb blockade occurs, so that an electron of spin $\sigma$ cannot tunnel into/out of the dot after an electron of opposite spin $-\sigma$ has already tunneled in/out. 
In other words, the particle number decay then coincides with the first tunneling process -- from an either empty or doubly occupied dot -- in which all single-particle states are available for transport. This causes the rate to approach~\cite{Splettstoesser10} the maximal value $\gamma_c \approx \Gamma_\uparrow + \Gamma_\downarrow = \Gamma$.\par

For fixed nonzero bias, the charge relaxation rate still equals $\Gamma/2$ when the stationary occupation is zero ($\epsilon-\mu>V/2$) or two ($\epsilon-\mu<-U-V/2$). For the intermediate $\epsilon$-regimes, there are now two cases:\par

(i) For $0 < V < U$, the singly-occupied state persists in a more narrow Coulomb-blockaded regime $-(U-V/2)<\epsilon-\mu<-V/2$. Here the enhancement of the charge rate to $\gamma_c=\Gamma$ can still be observed (dashed line in \Fig{fig::Decay_Rates}). In an $\epsilon$-window of width $V$, opened up on either side of this regime, single-electron processes cause left- and right decay channels to differ. This results in an new intermediate value for the charge relaxation rate, $\gamma_c=3\Gamma/4$ (shoulder of the dashed curve and also green dashed-dotted curve).\par

(ii) For $V>U$, \emph{both} transition energies can be in the bias window, giving rise to a new regime for which $\gamma_c\approx\Gamma/2$ (green dashed-dotted line in \Fig{fig::Decay_Rates}). Once the bias exceeds the charging energy, the Coulomb blockade is lifted for any level position $\epsilon$.\par

Finally, the spin rate $\gamma_s$ in \Fig{fig::Decay_Rates} (blue) behaves complementary to the charge rate $\gamma_c$, being suppressed relative to $\Gamma/2$ when the latter is enhanced and \emph{vice-versa}. When charge can (not) decay into lead $r$ through different spin channels, \emph{average} spin-flip processes are prohibited (possible)~\cite{Splettstoesser10,Contreras12}. Quantitatively, this is dictated by a sum rule for these two rates in our model, $\gamma_c +\gamma_s=\Gamma$, found in Ref.~\cite{Saptsov12a} and discussed below \BrackEq{eq::quantum_dot::rates::sum_rule}.

\paragraph{Time-dependent currents}

With the relaxation rates at hand, we can now analyze the effect of the bias $V$ on the time-dependent currents \eq{eq::quantum_dot::time_dependent_currents} in response to a parameter switch ($\rho_\In \neq z$). In particular, we consider a switch in which we abruptly change the level position from $\epsilon_\In$ to $\epsilon$ by the gate voltage. For the initial state $\Ket{\rho_\In}$, we therefore take the stationary state for level position $\epsilon_0$, keeping all other parameters the same [i.e., $\rho_\In(\epsilon_\In)=z(\epsilon_\In)$]. Written in the form \eq{eq::quantum_dot::stationary_state}, $\Ket{\rho_\In}$ is specified by just two parameters, $N_\In=\Braket{N}{\rho_\In}$ and $p_\In=\Braket{\fpop}{\rho_\In}$ since we here assume a non-spinpolarized initial state. In this situation, the only experimental probes are the charge and heat current.\par

The charge current out of the dot into lead $r$ is proportional to the charge decay rate for lead $r$ only, $\gamma_{cr}$:
\begin{align}
I^r_N(t) &= -\Bra{N}W \Ket{\rho(t)} = \gamma_{cr}\Big[ \Braket{N}{\rho(t)} - N_{z_r} \Big] \notag\\ 
&= \gamma_{cr}\Big[\left(N_{0}-N_z\right) \cdot e^{-\gamma_c t} - \left(N_{z_r}-N_z\right) \Big]. \label{eq::quantum_dot::particle_current_leadr}
\end{align}
Apart from this, it is proportional to the time-dependent excess occupation of the dot $\Braket{N}{\rho(t)}$ relative to $N_{z_r}=\Braket{N}{z_r}$, the stationary occupation that the dot would have if it were coupled \emph{only} to lead $r$. This deviation can be due to either the parameter switch, the first term\footnote{Although we consider the current at a specific junction $r$, this term decays with the full charge decay rate $\gamma_c=\sum_r \gamma_{cr}$, since this decay process can be accommodated by all leads. This becomes important for sufficiently large bias, see above.} in \Eq{eq::quantum_dot::particle_current_leadr} [since $N_\In=N_z$ if $\rho_\In=z$] or due the bias, the second term [since $N_{z_r}=N_z$ if $V=0$]~\footnote{$V=0$ implies $W_r \propto W$, giving $\Ket{z_r}=\Ket{z}$ for eigenvalue $0$ and $\Ket{p_r}=\Ket{p}$ with different eigenvalues $\Gamma_r$ and $\Gamma$.}.\par

The heat current out of the system into lead $r$ is of particular importance, since it is a multi-particle observable: by the general arguments of \Sec{sec::general::fermion_parity_mode}, it is sensitive to the fermion-parity mode~\cite{Schulenborg16,SchulenborgLicentiate}. It is given by~\cite{Schulenborg16} the flow of excess energy out of the dot relative to the electrochemical potential $\mu_r$ of  the receiving reservoir,
with the final result:
\begin{gather}
I_Q^r(t) = -\Bra{H-\mu_r N}W_r\Ket{\rho(t)}\label{eq::quantum_dot::heat_current_leadr}\\
= \Big[ \epsilon-\mu_r+\frac{U}{2}\left(2-\dual{N}_{z_r}\right) \Big] I_N^r(t) + \gamma_{pr}U\Braket{{p}_r'}{\rho(t)}.\notag
\end{gather}
The first part of the heat current \eq{eq::quantum_dot::heat_current_leadr} is thus directly proportional to the charge current and a characteristic energy. This energy (in square brackets) has a nontrivial parameter dependence which enters through the occupation $\dual{N}_{z_r}=\Braket{N}{\dual{z}_r}$ in the stationary state $\dual{z}_r$ achieved in the \emph{dual system} if it were only coupled to lead $r$ [\Eq{eq::quantum_dot::time_dependent_currents} ff.]. The second part of the heat current \eq{eq::quantum_dot::heat_current_leadr} due to the parity mode is associated to the dissipation of the Coulomb interaction energy: it equals $\Gamma_r U$ times the factor
\begin{subequations}
\begin{align}
\Braket{p'_r}{\rho(t)} &= \Braket{p'}{\rho_0}e^{-\gamma_p t}\label{eq:overlapb}\\
& + \Braket{p'_r}{z} + \frac{1}{2} \left(N_{0}-N_z\right)\left(\dual{N}_{z_r}-\dual{N}_z\right)e^{-\gamma_c t}. \label{eq:overlapc}
\end{align}
\label{eq::parity_contribution}
\end{subequations}
The term \eq{eq:overlapb} is a purely time-dependent effect of the switch, as it vanishes for $\rho_\In=z$ by orthogonality of $W$'s eigenvectors: $\Braket{p'}{z}=0$. In the zero bias case $V=0$, this term is the only non-zero contribution to \Eq{eq::parity_contribution}, as $\dual{z}_r \rightarrow \dual{z}$ and thus $\Braket{p'_r}{z} = \Braket{p'}{z} = 0 = \dual{N}_{z_r}-\dual{N}_z$. Hence, it was argued~\cite{Schulenborg16} to make the heat current a convenient observable to detect the fermion-parity rate $\gamma_p$.\par

For $V \neq 0$, the first additional term in \eq{eq:overlapc} gives a time-independent, purely bias related effect, while the second yields an additional $\gamma_c$-decay term. Most important for us is, however, that the $\gamma_p$-decay contribution in \eq{eq:overlapb} is also modified by the bias, $I_Q^r = a_p^r e^{-\gamma_p t} + \ldots $\,\,. In the following, we investigate whether this bias-induced modification of the amplitude
\begin{equation}
  a_p^r = \Gamma_{r}U \left[ \frac{1}{2}\left(\dual{N}_z-1\right) \left(N_{0}-1\right) + \frac{1}{4}\left(\dual{p}_z+p_{0}\right)\right]\label{eq::ap}
\end{equation}
alters or even spoils the previously proposed~\cite{Schulenborg16} detection scheme of $\gamma_p$ based on the heat current. The analysis of this problem is again substantially simplified by the duality: since $N_\In$, $p_\In$ depend on $\epsilon_\In$ in the same way as $N_z$ and $p_z$ depend on $\epsilon$, we only need to consider the behavior of the charge and the parity for the two stationary-state problems, namely for the actual system ($N_z$, $p_z$) and for the dual system ($\dual{N}_z$, $\dual{p}_z$).

\paragraph{Stationary occupations $N_z$ and $\dual{N}_z$}

\begin{figure}[t]
\includegraphics[width=1.0\columnwidth]{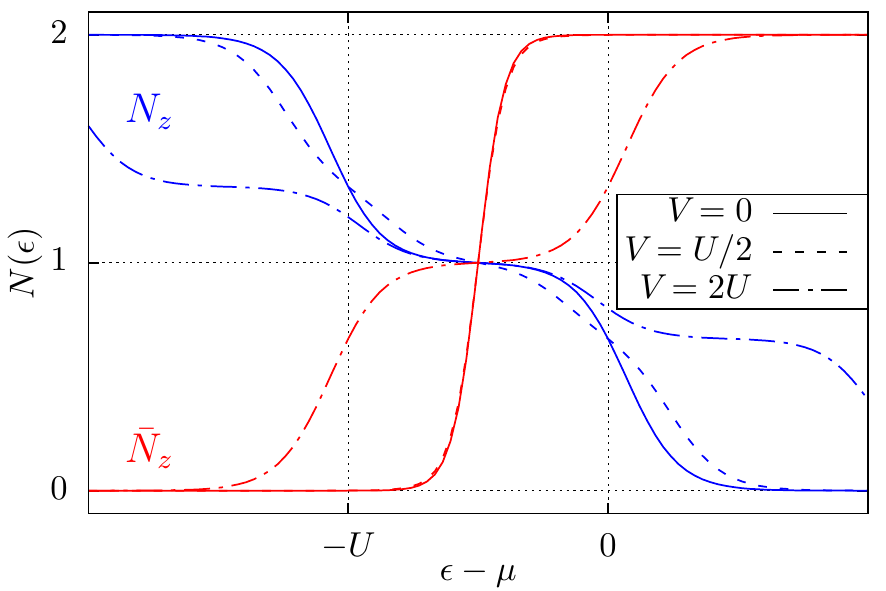}
\caption{Stationary occupation numbers of the actual quantum dot, $N_z$ (blue), and for the dual model with inverted energies, $\dual{N}_z$ (red). Both are plotted a function of the relative dot energy level $\epsilon-\mu$ for three bias values using the same conventions and parameters as in \Fig{fig::Decay_Rates}.}
\label{fig::occupations}
\end{figure}

In \Fig{fig::occupations}, we compare $N_z$ and $\dual{N}_z$ as a function of the dot level $\epsilon-\mu$. We first discuss moderate bias voltages $V\lesssim U$. For the actual  model, as is well known, the stationary occupation $N_z$  shows step-wise changes only when crossing the two Coulomb-resonances $\epsilon-\mu_r=0,-U$. In contrast, the dual system is doubly occupied (empty) when the actual dot model is empty (doubly occupied), as \Fig{fig::quantum_dot::model_dualmodel}(a) and (c) explain. While the actual dot can also be singly occupied, the attractive interaction in the dual system prohibits this for $T \ll U$: a singly-occupied, attractive quantum dot is unstable since it immediately either attracts an additional electron or emits an already present electron~\cite{Anderson75,Haldane77,Taraphder91}. 
Characteristic of this ``negative-$U$'' Coulomb blockade in the dual system is that this holds for \emph{any level position} $\epsilon$ at fixed $V<U$, i.e., even close to $\epsilon-\mu_r=0,-U$ where the Coulomb blockade in the actual model is already lifted for $V=0$. Hence, the dual occupation is either $\dual{N}_z = 0$ or $\dual{N}_z = 2$, with a sharp transition at the particle-hole symmetric point $\epsilon-\mu=-U/2$ typical for the attractive model~\cite{Anderson75}. One should note that the value $\dual{N}_z=1$ at this transition is only due to statistical mixing of $\Ket{0}$ and $\Ket{2}$ and \emph{not} because $\Ket{1}$ has any sizable weight, see \Fig{fig::quantum_dot::model_dualmodel}(b).\par

It is only at large bias $V>U$ that the Coulomb blockade in the dual system is lifted, so that the dual stationary state $\Ket{\dual{z}}$ can become a statistical mixtures of all four states $\Ket{0}$, $\KetUp$, $\KetDown$ and $\Ket{2}$. This results in a plateau where $N_z = \dual{N}_z = 1$ (red dashed-dotted line in \Fig{fig::occupations}): the step-wise change of $\dual{N}_z$ is no longer located at a single point but at two resonances. This bifurcation of the step positions of the dual occupation $\dual{N}_z$ as a function of the bias $V$ is the tell-tale sign of a lifted Coulomb-blockade in the \emph{dual model} due to non-equilibrium transport processes.

\paragraph{Parity contribution to the heat current} 

\begin{figure}[t]
\includegraphics[width=1.0\columnwidth]{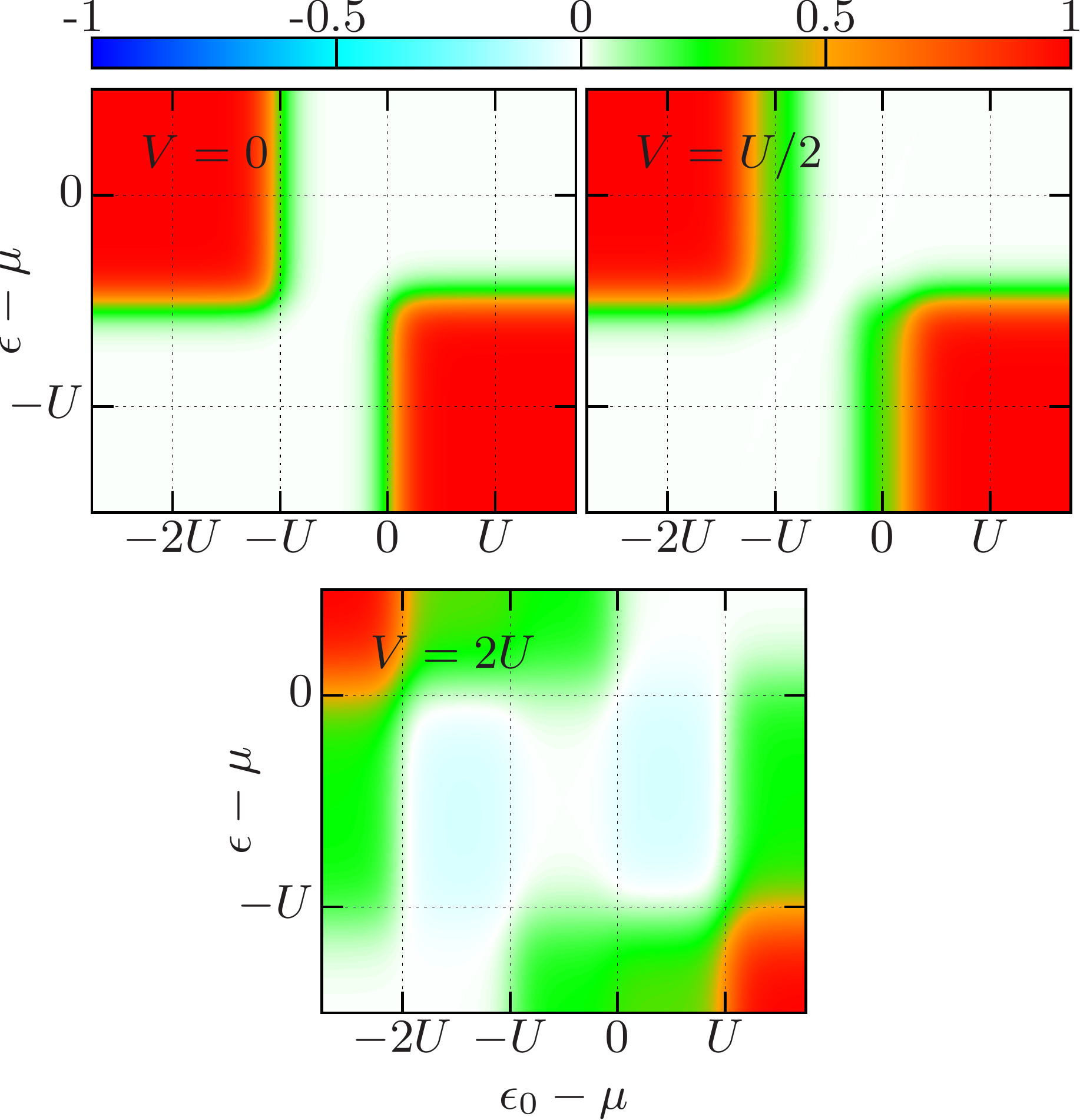}
\caption{Parity contribution to the heat current 
($I_Q^r= a_p^r e^{-\Gamma t} + \ldots$).
Color plot of the parity amplitude, $a_p^r$ in units of $\Gamma_{r} U$
as a function of the level position before ($\epsilon_\In-\mu$) and after ($\epsilon-\mu$) the switch for three values of the bias.
Same parameters as in \Fig{fig::Decay_Rates} and \fig{fig::occupations}.
\label{fig::ParityAmplitude}}
\end{figure}

With these dual physical pictures in mind, it is now possible to understand in detail the effect of the bias on the time-dependent heat current after the switch, in particular the amplitude $a_p^r$ of the fermion-parity mode [\Eq{eq::ap}]. In \Fig{fig::ParityAmplitude}, we compare the amplitude for several bias voltages, plotted as a function of the initial and final level of the switch $\epsilon_\In \rightarrow \epsilon$.\par

For $V=0$, the leftmost panel of \Fig{fig::ParityAmplitude}, we find a result analogous to that for the single-lead case studied in \Ref{Schulenborg16,SchulenborgLicentiate}. It has the striking feature that the amplitude as a function of the final level $\epsilon$ shows only a step-wise change at the electron-hole symmetry point, and \emph{not} at the expected Coulomb blockade resonances, as for the $\epsilon_\In$ dependence. This can be directly attributed to the attractive dual system discussed above: all $\epsilon$-dependence of $a_p^r$ enters through dual-system quantities $\dual{N}_z$ and $\dual{p}_z$, whereas the $\epsilon_0$-dependence follows the behavior of the actual model, $N_z$ and $p_z$.\par

An important conclusion of this work is that this behavior, and hence switch protocols suitable to detect the parity rate, remain basically unaltered when a substantial bias $V\lesssim U$ is applied, as can be seen from the right panel of \Fig{fig::ParityAmplitude}. It is only for the case of large bias $V>U$, shown in the central panel of \Fig{fig::ParityAmplitude}, that the properties of the amplitude $a_p^r$ change drastically. Again, this is directly related to the behavior of $\dual{N}_z$ and $N_z$ that we have analyzed in \Fig{fig::occupations}. Now the features of $a_p^r$ as a function of $\epsilon_\In$ occur at $-V/2-U\ (=-2U\ \text{for }V=2U)$, at $-V/2,\ V/2-U$ and $V/2$. In contrast, the features of $a_p^r$ as a function of the final level position can be found at $0$ and $-U$. This reflects the bifurcation of the $\epsilon$-positions of the steps in $\dual{N}_z$. 
The drastic change of the properties of the dual model due to the lifted Coulomb blockade thus directly impacts the bias dependence of the time-dependent heat current.

\subsection{Finite magnetic field}
\label{sec::quantum_dot::magnetic_field}
We conclude our study by discussing the decay dynamics in the presence of both a finite and possibly large magnetic field $B$ in addition to a finite bias voltage $V$. Also, the initial state $\rho_\In$ before the switch is now allowed to be spin polarized, $S_\In :=\Braket{S}{\rho_\In} \neq 0$.

\paragraph{Symmetry properties of decay rates}

\begin{figure}[t]
\centering
\includegraphics[width=1.0\columnwidth]{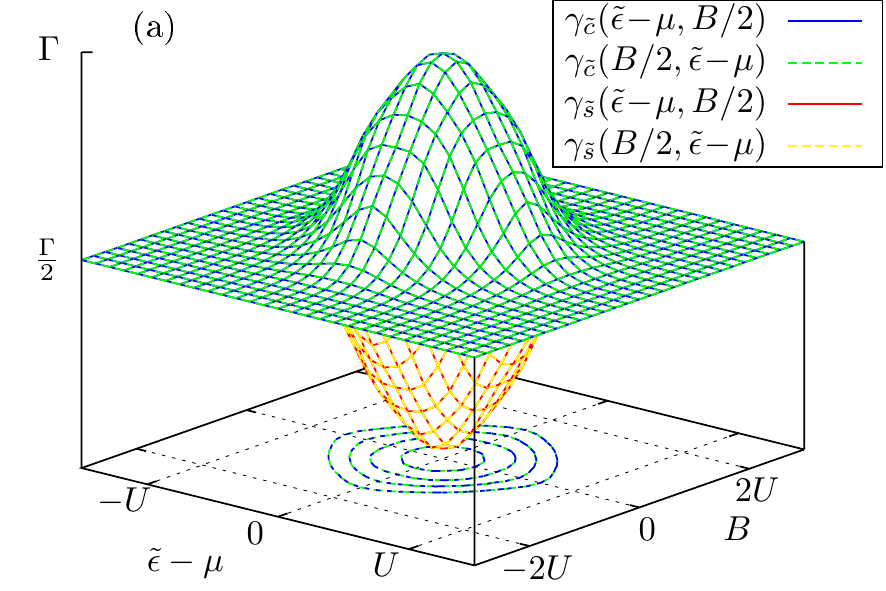}\\
\vspace{0.1cm}
\includegraphics[width=1.0\columnwidth]{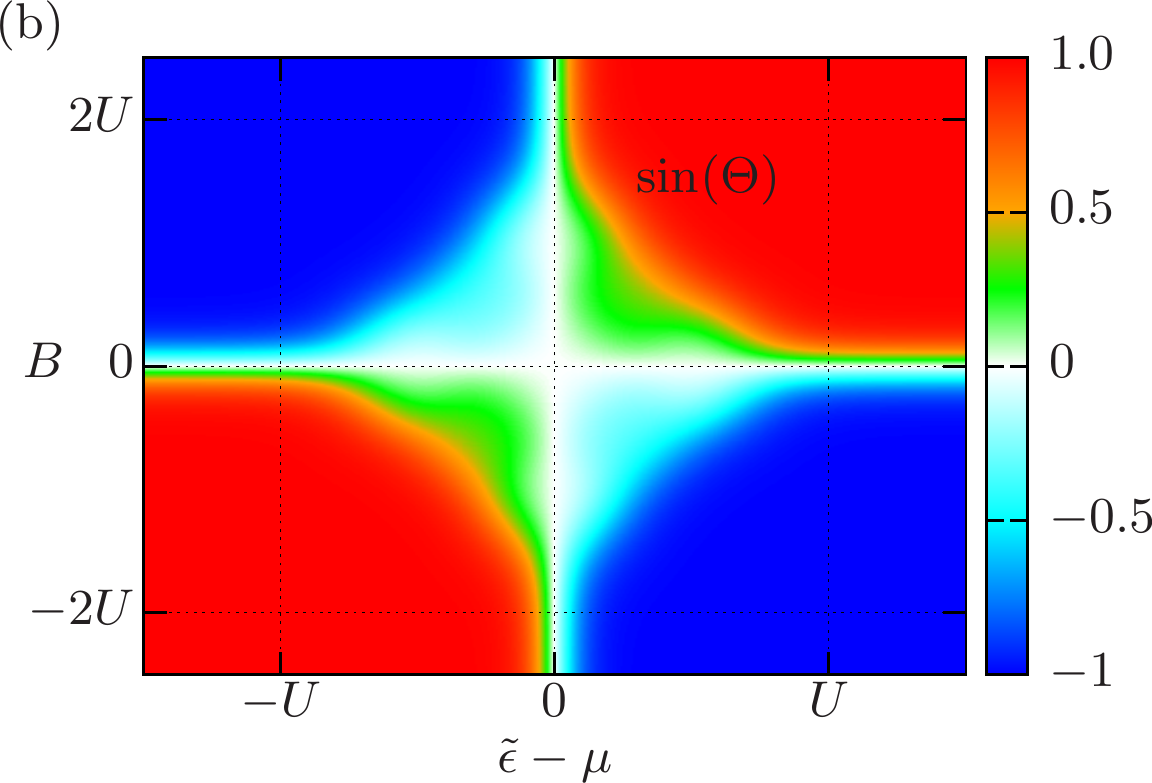}
\caption{Finite magnetic field:
(a)~Charge- ($\gamma_{\tilde{c}}$, blue and green) and spin-relaxation ($\gamma_{\tilde{s}}$, red and yellow) rate as a function of centered level $\tilde{\epsilon}$ and magnetic field $B$. In the $(\tilde{\epsilon},B)$ plane contour lines are drawn for $\gamma_{\tilde{c}}=0.6\Gamma, 0.7\Gamma, 0.8\Gamma, 0.9\Gamma$. (b)~Mode-mixing coefficient $\sin(\ThetaC)$ in the currents \BrackEq{eq::quantum_dot::charge_and_spin_current_magnetic} as a function of $\tilde{\epsilon}$ and $B$. In both (a) and (b), the bias is fixed at $V=U/2$. All other parameters are the same as in \Figs{fig::Decay_Rates}-\fig{fig::ParityAmplitude}.}
\label{fig::quantum_dot::magnetic_symmetry}
\end{figure}

We focus on the exact decay rates $\gamTilC$ and $\gamTilS$ \BrackEq{eq::quantum_dot::relaxation_rates}. In \Fig{fig::quantum_dot::magnetic_symmetry}(a), we plot the dependence of these rates on the level position and the magnetic field $B$. Before discussing how these rates result from charge- and spin-mode mixing -- requiring explicit computation -- we note a number of striking symmetries of this result that are of more general nature.\par

First of all, the well-known particle-hole symmetry of the system with respect to $\epsilon - \mu = -U/2$ and $B = 0$ manifests itself in an invariance under a simultaneous inversion of the signs of the magnetic field $B$, the ``centered'' level-position $\tilde{\epsilon} = \epsilon + U/2$, \emph{and} the bias $V = \mu_\mathrm{L} - \mu_\mathrm{R}$:
\begin{equation}
 \gamma_{\tilde{c}/\tilde{s}}(\tilde{\epsilon} - \mu,B,V) = \gamma_{\tilde{c}/\tilde{s}}(-(\tilde{\epsilon} - \mu),- B,-V),\label{eq::quantum_dot::ph_symmetry}
\end{equation}
which can be proven explicitly using \Eq{eq::quantum_dot::relaxation_rates} and \Eq{eq::quantum_dot::coupling_constant}. Second, a general analysis of the Anderson model has shown\footnote{It follows from Eqs. 134,140,155 and 232 of \Ref{Saptsov12a} (see also Eq. 112a of \Ref{Saptsov14a}) taken in the linear order in $\Gamma_{r\sigma}$.} that all the relaxation rates appear in pairs, each pair summing up to the parity rate,
\begin{equation}
 \gamTilC + \gamTilS = \gamma_p=\Gamma\label{eq::quantum_dot::rates::sum_rule}
\end{equation}
which is independent of $\tilde{\epsilon}$, $B$, $U$ and the ratios of all rates $\Gamma_{r\sigma}$. Clearly, the two surfaces plotted in \Fig{fig::quantum_dot::magnetic_symmetry}(a) obey this symmetry. There is, however, a third symmetry that is evident from \Fig{fig::quantum_dot::magnetic_symmetry}(a), but was not noted before: for symmetric couplings to the leads, $\Gamma_{\mathrm{L}\sigma} = \Gamma_{\mathrm{R}\sigma}$, the rates are also invariant when swapping the \emph{centered} level-position $\tilde{\epsilon}$ and its magnetic field \emph{shift} $B/2$:
\begin{equation}
 \gamma_{\tilde{c}/\tilde{s}}(\tilde{\epsilon} - \mu,B/2,V) = \gamma_{\tilde{c}/\tilde{s}}(B/2,\tilde{\epsilon} - \mu,V).\label{eq::quantum_dot::epsilon_B_exchange}
\end{equation}
This relation is a nontrivial consequence of our general duality \eq{eq::general::duality_relation}, the sum rule \eq{eq::quantum_dot::rates::sum_rule}, and a known duality for the Anderson model due to Iche~\cite{Iche72,Taraphder91,JensKoch}.\par

We stress that the duality of Iche is not equivalent to our duality \eq{eq::quantum_dot::duality_relation}, even though both involve a sign-inversion of energies, $(\epsilon, B, U) \to (-\epsilon, -B, -U)$. There are several reasons for this -- see \App{App::Modified_duality} -- but it is demonstrated most clearly by the fact that the composition of the two duality transformations is not the identity, and instead gives a new nontrivial duality: in \App{App::Modified_duality} we find for the Born-Markov kernel with symmetric couplings $\Gamma_{L\sigma} = \Gamma_{R\sigma}$ that
\begin{equation}
 W^\sudag(\tilde{\epsilon} - \mu,B/2,U,\mu_r) = -\Gamma - \tilde{\mathcal{P}}W(B/2,\tilde{\epsilon} - \mu,U,\mu_r)\tilde{\mathcal{P}}.\label{eq::quantum_dot::modified_duality}
\end{equation}
This \emph{swaps} the role of the gate-voltage ($\tilde{\epsilon} - \mu$) and the magnetic field ($B/2$), \emph{without} affecting the bias. The superoperator $\tilde{\P}$ appearing here is a combination of the parity superoperator $\P$ in our duality \eq{eq::quantum_dot::duality_relation} and a further transformation given in \App{App::Modified_duality}. This establishes a \emph{cross}link between the charge and spin decay rates,
\begin{equation}
  \gamma_{\tilde{c}/\tilde{s}}(\tilde{\epsilon} - \mu,B/2,U,\mu_r) = \Gamma - \gamma_{\tilde{s}/\tilde{c}}(B/2,\tilde{\epsilon} - \mu,U,\mu_r),
\end{equation}
whereas our relation \eq{eq::self-dual} expresses a \emph{self}-duality of the decay rates. Together with the sum rule \eq{eq::quantum_dot::rates::sum_rule}, this implies the new symmetry \eq{eq::quantum_dot::epsilon_B_exchange}. The duality \eq{eq::quantum_dot::modified_duality} constitutes a new aspect relative to previous works~\cite{Iche72,Taraphder91,JensKoch} by combining two nontrivial dualities and by extending the analysis to the relaxation \emph{dynamics}.\par

The new symmetry \eq{eq::quantum_dot::epsilon_B_exchange} in the parameter dependence of the relaxation properties of a single-level quantum dot has some interesting, practical consequences. If experimentally magnetic fields $B \sim U$ can be achieved, then mapping out the full $(B,\tilde{\epsilon})$ dependence on the scale $U$ to verify \Eq{eq::quantum_dot::epsilon_B_exchange} would constitute a stringent experimental test of the duality relation discussed in this article. On the other hand, whenever our model applies and such high fields are hard to maintain or control experimentally, then the symmetry \eq{eq::quantum_dot::epsilon_B_exchange} allows to obtain information about relaxation at high magnetic fields by studying the decay rates for \emph{small} field $B \ll U$ and large centered level position $\tilde{\epsilon} \sim U$. Tuning $\tilde{\epsilon}$  by a gate voltage is typically possible over much larger energy ranges than for $B$.

\paragraph{Mixing of charge- and spin-decay modes}

We now analyze in more detail how the exact rates $\gamTilC$ and $\gamTilS$ arise. Compared to the zero-field relaxation, the magnetic field introduces a qualitatively new feature by coupling the dynamics of the \emph{average} dot spin and charge, as mentioned after \Eq{eq::quantum_dot::relaxation_rates}. Also in this case, the intuition what mode-mixing entails in open systems is quite different from that in closed systems where the magnetic field only mixes spin states. To clarify this, we consider spin-independent tunneling, $\Gamma_{r\uparrow} = \Gamma_{r\downarrow} = \Gamma_r/2$, and write the exact relaxation rates \eq{eq::quantum_dot::relaxation_rates} as
\begin{equation}
 \gamma_{\tilde{c}/\tilde{s}} = \cos^2(\ThetaC/2)\cdot\gamma^{\mix}_{c/s} + \sin^2(\ThetaC/2)\cdot\gamma^{\mix}_{s/c},\label{eq::quantum_dot::rates::equal_bare_coupling}
\end{equation}
with the corresponding amplitude covectors
\begin{equation}
 \begin{pmatrix} \BraTilC \\ \BraTilS \end{pmatrix} = \begin{pmatrix}\cos(\ThetaC/2) && \sin(\ThetaC/2) \\ \sin(\ThetaC/2) && \cos(\ThetaC/2) \end{pmatrix} \cdot \begin{pmatrix}\Bra{c'} \\ \Bra{s'} \end{pmatrix}. \label{eq::quantum_dot::amplitudes_simple}
\end{equation}
By duality, there is a corresponding formula for the mixing of $\Ket{c}$, $\Ket{s}$ to form the exact modes $\Ket{\tilde{c}}$ and $\Ket{\tilde{s}}$. Although the matrix in \Eq{eq::quantum_dot::amplitudes_simple} does not represent a rotation, we denote $\ThetaC \in (-\pi/2,\pi/2)$ as the ``angle''  quantifying the mode-mixing, defined by $\sin(\ThetaC) := {2\DelGamS}/({\gamMixC - \gamMixS}) = {2\DelGamC}/({\gamMixS - \gamMixC})$. In the case of spin-independent tunneling, the single parameter $\ThetaC$ entirely captures the mode-mixing. Physically, it quantifies to which extent the transient average charge and spin currents decay independently. The interdependence of these currents generally stems from the fact that they are determined by the occupation probabilities of all states, which have both charge and spin quantum numbers.
\footnote{Only for $B=0$, one can use conservation laws to effectively ``average'' over the spin-quantum numbers and explicitly decouple the charge and spin dynamics, as shown in \Ref{Schulenborg16}.} For example, the total transient charge and spin currents flowing out of the dot are a mixture of the instantaneous deviations from the stationary value [\Eq{eq::quantum_dot::parity_covector}] of the charge, $\Delta N(t) = \Braket{c'}{\rho(t)}$, and the spin, $\Delta S(t) = \tfrac{1}{2} \Braket{s'}{\rho(t)}$ [cf. \Eq{eq::quantum_dot::charge_and_spin_amplitude}]:
\begin{align}
 I_N(t) &=  \gamMixC\Delta N(t) +\sin(\ThetaC) \cdot (\gamMixC - \gamMixS)\cdot\Delta S(t),\notag\\
 I_S(t) &=  \gamMixS\Delta S(t) + \sin(\ThetaC) \cdot (\gamMixS - \gamMixC)\tfrac{1}{4}\cdot\Delta N(t).\label{eq::quantum_dot::charge_and_spin_current_magnetic}
\end{align}
For $B = 0$, one finds the angle $\ThetaC = 0$, and for $U = 0$, we have $\gamma_c = \gamma_s$, so that in both cases, \Eq{eq::quantum_dot::charge_and_spin_current_magnetic} predicts charge and spin to decay independently with equal rate. For spin-independent tunneling, the charge- and spin-mode mixing is thus intimately related to the \emph{simultaneous} presence of a magnetic field and strong Coulomb correlations. In this case, the mixing angle can approach its extremal values $\ThetaC \rightarrow \pm \pi/2$ while $\gamma_c \neq \gamma_s$. This happens whenever $|B|$ and $\tilde{\epsilon} - \mu$ are such that the decay of the occupation of \emph{only one} spin state $\sigma$ is independent of the presence of another electron, whereas the occupation with an opposite spin $-\sigma$ \emph{does} depend on whether there is a second electron or not.
\footnote{More precisely, given $\sigma B > 0$ with $\sigma = \uparrow,\downarrow = +,-$ and $\Theta \rightarrow \sigma\pi/2$, only the spin $\sigma$ occupation does in general decouple and decay at a single rate, $\Delta N_{\sigma}(t) \propto e^{-(\Gamma/2) t}$. This can be seen by taking the limit $\Theta \rightarrow \pi/2$ in \Eq{eq::quantum_dot::rates::equal_bare_coupling} and \Eq{eq::quantum_dot::amplitudes_simple}, realizing that the latter predicts $\frac{1}{\sqrt{2}}[\Bra{c'} + \sigma\Bra{s'}] \sim \Bra{\Delta N_\sigma}$ to become a left eigenvector. The other spin occupation $N_{-\sigma}$ does \emph{not} correspond to a decay eigenmode, but to a nontrivial many-body excitation.}\par

In \Fig{fig::quantum_dot::magnetic_symmetry}(b), we plot the mode-mixing coefficient $\sin(\ThetaC)$ in \Eq{eq::quantum_dot::charge_and_spin_current_magnetic} as a function of the centered level position $\tilde{\epsilon}=\epsilon+U/2$ and the magnetic field $B$. As discussed, spin and charge always relax independently on the line $B = 0$ in \Fig{fig::quantum_dot::magnetic_symmetry}(b) where $\ThetaC = 0$. This can also be considered a result of full SU(2) symmetry of spin-rotations~\cite{Schulenborg16}, dictating the spin to decouple from the rest. However, in \Fig{fig::quantum_dot::magnetic_symmetry}(b), this also happens on the line $\tilde{\epsilon}-\mu=0$ ($\epsilon - \mu = -U/2$). This can be considered a result of full SU(2) symmetry of charge-rotations~\cite{Saptsov12a} (particle-hole symmetry), which dictates the charge to decouple from the rest. 
This is accounted for by the third, new symmetry \eq{eq::quantum_dot::epsilon_B_exchange} which essentially trades the gate voltage $\sim \tilde{\epsilon}-\mu$ coupling to charge for the magnetic field $B$ coupling to spin. However, \Fig{fig::quantum_dot::magnetic_symmetry}(b) mainly consists of red and blue regions in which $\ThetaC \rightarrow |\pi/2|$, indicating that for $\sigma\Theta > 0$, only the spin $\sigma$ occupation relaxes independently. Indeed, here both transition energies $\epsilon_\sigma - \mu$ and $\epsilon_\sigma - \mu + U$ of at least the spin value $\sigma$ fall outside the bias window. Interestingly, there are also well-defined narrow regimes (green/light blue) in which there is partial mode mixing, i.e., $\ThetaC \sim |\pi/4|$. This indicates that one spin occupation is independent of the other spin occupation \emph{only with respect to one of the two leads}.

\section{Outlook\label{sec::outlook}}
We have reviewed and extended the application of the new duality relation \eq{eq::general::duality_relation} that offers deep insight into the relaxation dynamics of quantum dots. Since several studies~\cite{Contreras12,Saptsov12a,Saptsov14a,Schulenborg14a,Schulenborg16,SchulenborgLicentiate} -- including the present one -- have established the importance of this unexplored fundamental aspect of relaxation properties of quantum dot systems, we believe that targeted experimental studies are now warranted. The applicability of these ideas to general classes of open electronic quantum systems allows several directions along which the experimental consequences can be worked out further. 
For example, while the concrete use of the duality has so far relied on the validity of the interacting single-orbital description, which is often good, the general physical insights we have gained should have similar implications for multi-level models, and should thus be analyzed by, e.g., extending~\Ref{Schulenborg14a}. Another simplification concerns the wide-band limit that we assumed. This is addressed in a forthcoming work~\cite{Schulenborg17} which demonstrates that even for strongly energy-dependent barrier transparencies, the duality has clear and practically useful consequences similar to those demonstrated here. Finally, we have combined our duality with the known Iche-duality to obtain a new symmetry of the parameter dependence of the relaxation rates. It is fundamentally interesting how this can be extended to more complex quantum dot systems along the above indicated lines.

\begin{acknowledgement}
We acknowledge the financial support of Erasmus Mundus (J. V.), DFG project SCHO 641/7-1 (R.B.S. and M.R.W), the Swedish VR (J.Sc., J. Sp.), and the Knut and Alice Wallenberg Foundation (J. Sp.). The authors thank F. Haupt and N. Dittmann for useful discussions on the topic.
\end{acknowledgement}

\appendix

\section{Kernel $W$ and stationary expectation values}\label{App::W_Expectations}
The kernel $W$ for the single-level quantum dot Hamiltonian \eq{eq::quantum_dot::hamiltonian} can be represented by the matrix
\begin{equation}
W=
\begin{bmatrix}
-\!\!\!\!\sum\limits_{\text{column}} & \Gamma_\uparrow f^-_{\epsilon\uparrow} & \Gamma_\downarrow f^-_{\epsilon\downarrow} & 0 \\
\Gamma_\uparrow f^+_{\epsilon\uparrow} & -\!\!\!\!\sum\limits_{\text{column}} & 0 & \Gamma_\downarrow f^-_{U\downarrow} \\
\Gamma_\downarrow f^+_{\epsilon\downarrow} & 0 & -\!\!\!\!\sum\limits_{\text{column}} & \Gamma_\uparrow f^-_{U\uparrow} \\
0 & \Gamma_\downarrow f^+_{U\downarrow} & \Gamma_\uparrow f^+_{U\uparrow} & -\!\!\!\!\sum\limits_{\text{column}}
\end{bmatrix},
\end{equation}
in the Liouville-space basis $\Ket{0}$, $\KetUp$, $\KetDown$ and $\Ket{2}$ defined in the main text after \Eq{eq::quantum_dot::reduced_density_operator}. By $\sum_{\text{column}}$, we indicate the sum over the other elements present in the same column, thereby guaranteeing probability conservation: $\Bra{\one}W = 0$. \par

The stationary state $\Ket{z}$ is determined by solving $W\Ket{z} = 0$ together with $\Braket{\one}{z} = 1$. By \Eq{eq::quantum_dot::stationary_state}, it can be parametrized by $N_z$, $p_z$ and $S_z$, which all involve the expression
\begin{align}
A^{-1}=&\left(f^+_{\epsilon \uparrow}+f^-_{U \uparrow}\right)\left(f^-_{\epsilon \downarrow}+f^+_{U \downarrow}\right)\notag\\
&+\left(f^-_{\epsilon \uparrow}+f^+_{U \uparrow}\right)\left(f^+_{\epsilon \downarrow}+f^-_{U \downarrow}\right)
\end{align}
containing the Fermi functions defined below \Eq{eq::quantum_dot::spin_resolved_rates}. The stationary particle number $N_z = \Braket{N}{z}$ is then explicitly given by
\begin{equation}
N_z = 2A\left[f^+_{\epsilon\uparrow}\left(f^+_{U\downarrow}+f^-_{\epsilon\downarrow}\right)+f^+_{\epsilon\downarrow}\left(f^+_{U\uparrow}+f^-_{\epsilon\uparrow}\right)\right],
\end{equation}
Similarly, the stationary spin $S_z = \Braket{S}{z}$ is found to be
\begin{equation}
S_z=2A\left(f^+_{\epsilon\uparrow}f^-_{U\downarrow}-f^+_{\epsilon\downarrow}f^-_{U\uparrow}\right),
\end{equation}
and the stationary fermion-parity $p_z = \Braket{\fpop}{z}$ reads
\begin{align}
p_z =&\frac{2A}{\Gamma}\left[\left(\Gamma_\downarrow f^-_{\epsilon\downarrow}+\Gamma_\uparrow f^+_{U\uparrow}\right)\left(f^+_{\epsilon\uparrow}f^+_{U\downarrow}+f^-_{\epsilon\uparrow}f^-_{U\downarrow}\right)\right.\notag \\
&+\left. \left(\Gamma_\uparrow f^-_{\epsilon\uparrow}+\Gamma_\downarrow f^+_{U\downarrow}\right)\left(f^+_{\epsilon\downarrow}f^+_{U\uparrow}+f^-_{\epsilon\downarrow}f^-_{U\uparrow}\right)\right.\notag \\
&- \left. \left(\Gamma_\uparrow f^-_{U\uparrow}+\Gamma_\downarrow f^-_{U\downarrow}\right)\left(f^+_{\epsilon\uparrow}f^-_{\epsilon\downarrow}+f^-_{\epsilon\uparrow}f^+_{\epsilon\downarrow}\right)  \right.\notag \\
&-\left. \left(\Gamma_\uparrow f^+_{\epsilon\uparrow}+\Gamma_\downarrow f^+_{\epsilon\downarrow}\right) \left( f^+_{U\uparrow}f^-_{U\downarrow}+f^-_{U\uparrow}f^+_{U\downarrow} \right) \right].
\end{align}

\section{Iche's duality [\Eq{eq::quantum_dot::modified_duality}]: derivation and comparison
\label{App::Modified_duality}}

\paragraph{Iche's duality mapping.} 

To derive \Eq{eq::quantum_dot::modified_duality}, we first rewrite the dot Hamiltonian \eq{eq::quantum_dot::hamiltonian} using $\fpop = 1-2N+4N_\uparrow N_\downarrow $ and $\tilde{\epsilon}:= \epsilon+U/2$ following~\cite{Saptsov12a}. Dropping a constant, we find
\begin{equation}
 H = \tilde{\epsilon}(N-1) + BS + \frac{U}{4}\fpop .\label{tildeepsH}
\end{equation}
Important for the following is that the first and second term are operators constructed from even [$N-1=\Ket{2}-\Ket{0}$] and odd particle number states [$S=\tfrac 1 2 \left[\KetUp-\KetDown\right]$], respectively, see the definitions after \Eq{eq::quantum_dot::reduced_density_operator}.\par

For the construction of a dual model in which all energy signs are inverted, the third term presents, however, a fundamental problem: a unitary operator acting only on the quantum dot that inverts all the energy signs in \Eq{tildeepsH} \emph{cannot} be built from physical observables. The reason is that any observable operator must be of bosonic type -- i.e., consisting only of even products of fields -- in order to obey the univalence postulate of quantum mechanics~\cite{Wick52}. By this fermion-parity superselection principle, \emph{any} observable (or mixed-state) operator must commute with the operator $\fpop$. No unitary built from such operators can thus change the sign of the third term on the right hand side of \Eq{tildeepsH}.\par

A transformation that inverts all signs must thus be of fermionic type, i.e., consisting only of products of an odd number of the field operators. For our quantum dot model, the simplest example of a fermionic operator that does this is~\cite{Iche72,Taraphder91} the Majorana fermion operator
\begin{equation}
 a_\uparrow = i(d_\uparrow^\dagger-d_\uparrow).
\end{equation}
Indeed, one can check that up to an irrelevant constant,
\begin{equation}
 a_\uparrow H(\tilde{\epsilon}, B/2, U)a_\uparrow = {H}(-B/2,-\tilde{\epsilon}, -U).\label{vonOppen-transform_local}
\end{equation}
Clearly, to achieve this, the operator $a_\uparrow$ must explicitly violate univalence: it maps the even parity subspace, spanned by $\Ket{0}$ and $\Ket{2}$, to the odd-parity subspace, spanned by $\KetUp$, $\KetDown$ and \emph{vice versa}. This thus necessitates the interchange of the role of $B$ and $\tilde{\epsilon}$ (apart from sign changes) on the right hand side of \Eq{vonOppen-transform_local}.\par

Transformation \eq{vonOppen-transform_local} applied to $H^\tun$ changes the form of the coupling. To restore it to the form of a tunnel Hamiltonian, one needs to apply a unitary to the leads as well, given by Koch et. al.~\cite{JensKoch}:
\begin{equation}
 \Xi=\exp \Big( \frac{\pi}{2}\sum\limits_k \Big[ c_{r k \uparrow}^\dagger c_{-{r}-k\uparrow}^\dagger-c_{-{r} -k \uparrow} c_{r k \uparrow} \Big] \Big),
\end{equation}
together with the parameter substitution operator $\Upsilon: \, t_{rk\uparrow} \rightarrow rt_{-{r} -k \uparrow}^*$, where $r=$L, R$=+,-$. Taken together, the unitary transformation $\Omega=a_\uparrow \Xi \Upsilon$ now maps the full model \eq{eq::general::hamiltonian_tot} with $H$ given by \Eq{tildeepsH} onto a model with the \emph{same} reservoir Hamiltonian $\widetilde{H}^\res=H^\res$ but with a modified quantum dot, $\widetilde{H}(\tilde{\epsilon},B/2,U) ={H}(-B/2,-\tilde{\epsilon}, -U)$, and tunnel coupling
\begin{equation}
 \widetilde{H}^\tun= \sum_{r,k,\sigma}\tilde{\tau}_{rk\sigma}c_{rk\sigma}^\dagger d_\sigma + \mathrm{H.c. },
\end{equation}
where $\tilde{\tau}_{rk\sigma}=r\tau_{-r-k\sigma}^* $, $\sigma=\uparrow$, and $\tilde{\tau}_{rk\sigma}={\tau}_{rk\sigma}$, $\sigma=\downarrow$. The mapping is essentially the one of \Ref{JensKoch}, except that we here used $a_\uparrow$ instead of $a_\downarrow$.

\paragraph{Difference to our duality \eq{eq::quantum_dot::duality_relation}.}

A key difference highlighted in our derivation of Iche's transformation is that it purposely breaks univalence locally whereas our duality respects it~\cite{Schulenborg16}. Furthermore, the above mapping requires perfectly symmetric bands~\cite{Taraphder91}, $\epsilon_{r -k\sigma}=-\epsilon_{r k \sigma}$, (which is implied by our wide-band limit) and a symmetric bias $\mu_r=-\mu_{-r}$ (which we also assume). Moreover, it has no formulation independent of the number of the reservoirs.\footnote{For example, for the case of an odd number of the reservoirs attached, the dual model reservoirs would also acquire magnetization: e.g., a single reservoir with a spin-degenerate quantum dot level will map to a model where the transition energies for spin $\uparrow$ and $\downarrow$ are different, i.e., there is an internal magnetization of the reservoirs in the dual model $\widetilde{H}^\res$.}
This makes the applicability of Iche's duality~\cite{JensKoch} more restricted as compared to our duality \eq{eq::general::duality_relation}. However, for the case of two symmetrically biased wide-band leads that we consider in the present paper, it turns out to be advantageous to combine these two dualities to obtain the new, nontrivial relation \eq{eq::quantum_dot::modified_duality}.

\paragraph{Iche's duality for time evolution kernels.}

Deriving the time evolution kernels along the lines of \Ref{Schoeller09a,Saptsov12a,Saptsov14a}, the above mapping yields a duality relation that does not involve the hermitian-adjoint. Introducing the Laplace frequency $z$, we obtain
\begin{equation}
  W(z)=\mathcal{Y}\, \mathcal{A}_\uparrow \widetilde{W}(z) \, \mathcal{A}_\uparrow \mathcal{Y}.\label{vonOppen_sigma}
\end{equation}
Here, $W(z)= W(z,\tilde{\epsilon},B/2,U, \mu_r)$ is the kernel for the actual model and $\widetilde{W}(z):= W(z,-B/2,-\tilde{\epsilon},-U,\mu_r)$ is the kernel for the Iche-dual system. Furthermore, $\mathcal{A}_\uparrow\bullet=a_\uparrow\bullet a_\uparrow$ is a superoperator generated by the Majorana field operator. Finally, $\mathcal{Y}$ denotes the parameter substitution generated by $\Upsilon$ which simply swaps the tunnel rates of the two leads, $\mathcal{Y} :\, \Gamma_{r\uparrow} \rightarrow \Gamma_{-r\uparrow}$, and can be omitted in the case of symmetric coupling. Taking~\cite{Schulenborg16} the Born-Markov limit [$z=i\eta \to i0$ together with $\Gamma \to 0$, cf. \Eq{eq::general::born_markov_equation}] and substituting the result into \Eq{eq::general::duality_relation}, we find \Eq{eq::quantum_dot::modified_duality} of the main text with $\widetilde{\mathcal{P}}=i\mathcal{Y}\mathcal{M}\mathcal{A}_\uparrow \mathcal{P}$.
The factor $i$ compensates the anticommutation sign of the superoperators $\mathcal{A}_\uparrow$ and $\mathcal{P}$. Finally, the parameter substitution $\mathcal{M}$: $\mu_r \rightarrow -\mu_r$ also inverts the bias. Note that for the case of symmetric couplings considered in the main text, this bias inversion does not change the decay rates, and can thus be dropped to arrive at the final symmetry relation \Eq{eq::quantum_dot::epsilon_B_exchange}.


\newcommand{\noop}[1]{}
\providecommand{\WileyBibTextsc}{}
\let\textsc\WileyBibTextsc
\providecommand{\othercit}{}
\providecommand{\jr}[1]{#1}
\providecommand{\etal}{~et~al.}

\end{document}